\documentclass[%
 reprint,
 amsmath,amssymb,
 aps,
 prx,
]{revtex4-2}
\usepackage{amsmath}
\usepackage{graphicx}% Include figure files
\usepackage{dcolumn}% Align table columns on decimal point
\usepackage{bm}% bold math
\usepackage{amsmath}
\usepackage{comment}
\usepackage{xcolor}
\usepackage{braket}
\usepackage{enumerate}

\newcommand{\dd}{\mathrm{d}}

\begin{document}

\preprint{APS/123-QED}

\title{Nonlocal Thermoelectric Transport in Degenerate Electron Systems}

\textbf{}
\author{Albert Beardo}
\email{albert.beardo@uab.cat}
\author{Adrián Ramos-Martín}
\author{Jordi Tur-Prats}

% \email{jordi.tur@uab.cat}

% \email{lluc.sendra@uab.cat}
%\author{Pol Torres}
%\author{Javier Bafaluy}
\author{F. Xavier Alvarez}
\author{Juan Camacho}
 %\email{juan.camacho@uab.cat}
\affiliation{Departament de Física, Universitat Autònoma de Barcelona, E-08193 Bellaterra, Catalonia, Spain}

\begin{abstract}
When the characteristic length scales of transport become comparable to the
carrier mean free path, the local constitutive description breaks down.
We develop the Flux Derivative Formalism (FDF), a kinetic framework that
provides a systematic route from the Boltzmann equation to nonlocal
constitutive equations. We use it to derive coupled equations for charge and heat transport in isotropic electronic systems, together with boundary conditions derived consistently within this approach, yielding a closed macroscopic description that can be applied to complex geometries without solving the full Boltzmann problem. The explicit transport coefficients are derived for degenerate electronic systems within the Callaway model. A central result is that varying the relative strength of
resistive and momentum-conserving collisions changes the transport
coefficients but not the non-local constitutive structure, providing a continuous
description from resistive-dominated transport to the collective
hydrodynamic limit. The electronic heat equation takes a Guyer--Krumhansl form, whose predictions are validated against size-dependent thermal-transport measurements in Cu and Ni. For coupled thermoelectric transport, the resulting boundary-value problem determines the complete geometry-dependent transport matrix. Comparison with spatially resolved Boltzmann solutions confirms quantitative accuracy down to transverse dimensions comparable to the carrier mean free path. The theory predicts geometry-dependent Seebeck and Peltier coefficients while preserving Onsager reciprocity and the Kelvin relation.
\end{abstract}

\maketitle

\section{Introduction}

Thermoelectric materials, which interconvert heat and electricity through the Seebeck and Peltier effects, are central to waste-heat energy harvesting and solid-state cooling technologies \cite{Snyder2008}. Their operation is described macroscopically by constitutive equations relating the electric and heat fluxes to the thermodynamic forces driving irreversible transport.
Under ordinary conditions, Ohm's law, Fourier's law, and the Seebeck--Peltier relations provide an accurate local description, in which the transport fluxes at a given position are completely determined by the local thermodynamic state and its gradients.

This local description breaks down when the characteristic transport length becomes comparable to the electronic mean free path. Under these conditions, the transport fluxes are influenced by the nonequilibrium carrier dynamics over finite spatial and temporal scales, giving rise to nonlocal transport. Constitutive equations containing higher-order spatial derivatives have consequently been proposed in a variety of contexts, including rarefied gases \cite{Struchtrup2005}, phonon heat transport \cite{Guo2015}, electron hydrodynamics \cite{Bandurin2016,Guo2017_Pnas,LucasFong2018,Narozhny2022,Fritz2024,HuiSkinner2025}, and mesoscopic electron systems \cite{varnavides2023charge,Baker2024}. Nonlocal electronic transport has in fact a long history, dating back to the classical treatments of size effects and the anomalous skin effect \cite{Fuchs1938,Sondheimer1952,Reuter1948}. Although these theories originate from different microscopic mechanisms, they all describe departures from locality associated with the finite propagation of nonequilibrium excitations.

The most striking recent manifestations of electronic nonlocality have emerged in charge transport. In ultraclean 2D materials such as graphene and tungsten ditelluride, spatially nonlocal effects in the charge current manifest as Poiseuille-like viscous flows and superballistic conduction---phenomena in which electron-electron interactions dominate over momentum-relaxing scattering, enabling collective hydrodynamic behavior. Direct imaging of Poiseuille flow profiles \cite{Sulpizio2019} and superballistic transport through graphene constrictions \cite{KrishnaKumar2017} provide compelling evidence of these nonlocal phenomena. Imaging of the viscous Dirac fluid in graphene \cite{Ku2020} further confirms that nonlocal constitutive equations describe observable physics with direct experimental signatures. These observations raise a broader question: how does electronic nonlocality manifest when charge and heat currents are coupled, and what signatures does it produce in thermoelectric transport? The present work systematically addresses this question by deriving nonlocal constitutive equations for coupled charge and heat transport.

A microscopic derivation of such constitutive equations naturally starts from the Boltzmann transport equation. However, obtaining a tractable macroscopic description requires reducing the large number of degrees of freedom contained in the distribution function to a finite set of nonequilibrium variables. This reduction is not unique: different kinetic closures, such as the Chapman--Enskog expansion \cite{Chapman1953} or Grad's moment method \cite{Grad1949}, prescribe different microscopic representations of the nonequilibrium distribution, and the resulting constitutive equations depend on the closure adopted.

In this work, we use the Flux Derivative Formalism (FDF) to construct the coupled electronic constitutive equations. The FDF is a reduced kinetic description in which the nonequilibrium distribution function is parameterized by the relevant transport fluxes and by their leading spatial and temporal derivatives. It therefore provides a systematic lowest-order extension of the local constitutive description capable of retaining nonlocal corrections while preserving a closed analytical formulation.
The FDF was originally developed for nonlocal phonon heat transport \cite{Sendra2021,Sendra2022}. In that context, a closure based on the heat flux and its first spatial and temporal derivatives was shown to reproduce the leading nonlocal corrections to Fourier transport. 
%Comparisons with numerical solutions of the phonon Boltzmann equation further indicated that the resulting constitutive equations remain quantitatively accurate up to Knudsen numbers of order unity \cite{Guo2018}. 
Importantly, these constitutive equations have been validated against experimental phonon transport data in silicon across diverse experimental conditions \cite{Alajlouni2020,Torres2018,Beardo2020,Beardo2021,beardo2022hydrodynamic}, establishing the robustness of this kinetic strategy for nonlocal transport. These results motivate its use to construct coupled electronic constitutive equations.

Related theoretical developments have explored nonlocal and high-order transport responses from the Boltzmann equation. Battiato, Zlati'c, and Held \cite{Battiato2017} derived transport coefficients depending on higher-order spatial derivatives of external fields. More recently, Coulter \emph{et al.} \cite{Coulter2025} developed viscous thermoelectric equations from coupled electron--phonon Boltzmann dynamics. The present approach differs from these formulations by retaining charge and heat currents as independent constitutive variables. In contrast to expansions in spatial derivatives of the driving forces, the nonlocality is expressed through derivatives of the physical currents; 
the resulting equations recover the hydrodynamic equations for collective electron transport not only in situations dominated by momentum-conserving collisions, but also maintain their constitutive structure as the relative strength of resistive scattering is varied.
Retaining both currents also preserves independent nonlocal charge and heat-current sectors, rather than reducing the closure to a single projected mode. %Together with kinetic boundary conditions formulated in terms of the same physical currents, the resulting equations define a closed boundary-value problem that determines the complete geometry-dependent transport matrix.

Starting from the semiclassical Boltzmann equation, we derive closed constitutive equations for coupled charge and heat transport in isotropic degenerate electronic systems. The constitutive structure follows from the kinetic closure, while the microscopic collision operator determines the transport coefficients. Onsager reciprocity provides an independent constraint on the resulting coupled equations and is satisfied to the retained Sommerfeld order for the collision model considered here. The equations contain two characteristic nonlocal length scales: a longer scale primarily associated with charge transport and a shorter scale primarily associated with heat transport. Together with boundary conditions derived consistently within this approach, they define a closed boundary-value problem directly in terms of the physical charge and heat currents. Its solution determines the complete geometry-dependent transport matrix.

The FDF equations recover the Stokes--Ohm description of viscous charge flow and a Guyer--Krumhansl-type equation for heat transport, thereby connecting the local limit, kinetic nonlocality, and collective transport within a common constitutive description. 
Remarkably, the resulting electronic heat-current equation quantitatively accounts for size-dependent thermal-transport measurements in Cu and Ni \cite{barako2016quasi,islam2024}, providing direct experimental support for a nonlocal constitutive description of electronic heat transport.
In the coupled thermoelectric problem, confinement modifies the Seebeck and Peltier coefficients while preserving the Kelvin relation. Comparison with spatially resolved Boltzmann solutions confirms quantitative accuracy down to transverse dimensions comparable to the carrier mean free path.

The remainder of this paper is organized as follows. We first introduce the
Flux Derivative Formalism as a kinetic closure of the Boltzmann equation and
apply it to isothermal charge transport, establishing its connection with the
Stokes--Ohm equation and the crossover between resistive and
momentum-conserving regimes. We then consider electronic heat transport; the
resulting heat-current equation has the Guyer--Krumhansl form and is confronted
with size-dependent thermal-transport measurements in Cu and Ni. We next
extend the construction to coupled charge and heat transport and derive the
corresponding nonlocal constitutive equations and their microscopic
coefficients within the Callaway model. \textcolor{black}{Their consequences for finite geometries are then analyzed in terms of the complete geometry-dependent transport matrix, with the Seebeck and Peltier coefficients providing experimentally relevant manifestations of the resulting finite-size response. The finite-width thermoelectric predictions are finally benchmarked against spatially resolved solutions of the Boltzmann equation.}

\section{Flux Derivative Formalism}\label{sec:FDF}

In this section we present the Flux Derivative Formalism, a systematic
procedure for deriving nonlocal constitutive equations and microscopic
expressions for their transport coefficients from the semiclassical Boltzmann
equation. We first introduce the kinetic closure underlying the procedure and
then derive the corresponding reduced kinetic description.

The derivation of constitutive equations from the Boltzmann transport equation
requires reducing the microscopic description contained in the distribution
function to a tractable set of macroscopic quantities. Such a reduction
necessarily requires a kinetic closure, namely, an additional assumption
allowing the nonequilibrium distribution function to be represented in terms of
a finite number of degrees of freedom. Different kinetic theories differ
primarily in the way this closure is introduced.

In the Chapman--Enskog method \cite{Chapman1953}, the closure is achieved through an asymptotic
expansion in the Knudsen number, leading to constitutive equations expressed in
terms of the gradients of the conserved thermodynamic variables. In Grad's
moment method \cite{Grad1949}, the nonequilibrium distribution is expanded in a finite basis of
microscopic moments, and the resulting hierarchy is closed by truncation.
Although conceptually different, both approaches prescribe a particular
microscopic representation of the nonequilibrium distribution before deriving
the associated macroscopic equations.

The present approach differs from conventional kinetic theories in the
choice of kinetic closure. As in Chapman--Enskog and Grad's methods, the
local-equilibrium distribution is completely determined by the conserved
thermodynamic variables. The difference concerns exclusively the construction
of the nonequilibrium correction. Instead of prescribing a \textit{microscopic 
representation} of the nonequilibrium distribution, the kinetic closure is 
defined by the choice of the \textit{macroscopic variables} retained in the reduced 
description. The corresponding microscopic coefficient functions are then 
determined from the Boltzmann equation.

For transport phenomena, the transport fluxes provide a natural description of
the nonequilibrium state, since they are precisely the variables entering the
local constitutive equations. This choice is also central to extended irreversible thermodynamics, where fluxes are treated as independent nonequilibrium variables \cite{Jou2010}. We therefore adopt the transport fluxes as the
fundamental nonequilibrium variables of the reduced kinetic description.
To account for weakly nonlocal transport, the nonequilibrium distribution is
allowed to depend additionally on the first spatial, and, when required,
temporal derivatives of the transport fluxes. This constitutes the
lowest-order extension beyond a purely local constitutive description.
Schematically, for a generic transport flux $\mathbf J$ this amounts to writing
\begin{equation}
f=f^0+
\vec{\beta}_{ J}\cdot\mathbf J+
\vec{\gamma}_{J}\cdot
\frac{\partial\mathbf J}{\partial t}
+\overset{=}{G}_{ J}:\nabla\mathbf J .
\end{equation}
where $\vec{\beta}_{J}$,
$\vec{\gamma}_{J}$, and
$\overset{=}{G}_{J}$ are the microscopic weight functions, determined from the Boltzmann equation once the retained fluxes are specified. The procedure to determine the transport equations and the weight functions from the Boltzmann equation is illustrated in the following section, where we apply it to specific transport regimes.

%The formulation of the FDF follows in five steps: (i) The nonequilibrium distribution is written as the constitutive expansion of Eq.~\eqref{ansatz_generic} in terms of the retained fluxes. (ii) The Boltzmann equation is projected onto microscopic quantities to obtain exact balance and constitutive equations, which still involve higher-order moments of the distribution function; these equations are closed by substituting the constitutive expansion, yielding closed transport equations for the retained fluxes. (iii) The constitutive expansion is substituted back into the full Boltzmann equation; the resulting explicit dependence on the thermodynamic forces ($\nabla T$, $\vec{E}$) is then eliminated using the balance equations obtained in step (ii). (iv) Terms involving second- and higher-order spatial and temporal derivatives are neglected, so that only the retained fluxes and their first derivatives survive, consistently with the constitutive expansion itself. (v) Identifying the coefficients of these independent variables provides the equations determining the weight functions.

The FDF approach is expected to remain accurate provided the
neglected higher-order spatial and temporal variations remain comparatively
small, i.e., when the transport length scales of interest are not much shorter than the electronic mean free path. More elaborate reduced
descriptions can be constructed systematically by enlarging the set of retained
macroscopic variables. 

%The alternative approach recently proposed by Coulter et al. \cite{Coulter2025} expands the distribution function in terms of microscopic electron-phonon ``relaxon'' excitations rather than the macroscopic transport variables adopted here.

In the next sections we apply the FDF method first to isothermal charge
transport, where $\mathbf{J}$ is the current density $\mathbf{j}$, then to heat transport, where
$\mathbf{J}$ is the heat flux $\mathbf{q}$, and finally to coupled thermoelectric transport,
where the retained fluxes are $\mathbf{j}$ and $\mathbf{q}$.

\section{Isothermal charge transport}\label{sec:charge_transport}

%We now apply the FDF to the simplest transport problem, namely isothermal charge transport. 
Isothermal charge transport illustrates the general methodology while keeping the formalism at its simplest, since the electric current is the only transport flux retained as an independent variable. Our objective is to derive its constitutive equation and the microscopic expressions for its coefficients directly from the Boltzmann transport equation. We first present the formalism for a general collision operator and then apply it to the Callaway model.

\subsection{General collision operator}
We now follow the systematic FDF procedure to derive the constitutive equation from the Boltzmann equation, with further details provided in Appendix~\ref{app:isothermal_derivation}. 

The linearized BTE for electrons writes: 
\begin{equation}
    \frac{\partial f}{\partial t}+\vec{v} \cdot {\nabla}_{\vec{r}} f + \vec{F} \cdot {\nabla}_{\vec{p}} f=C(f),\label{BTE}
\end{equation}
where $f(\vec{r},\vec{p},t)$ is the distribution function for electrons with momentum $\vec{p}=\hbar \vec{k}$ ($\vec{k}$ being the wave vector) in position $\vec{r}$ at time t, $\vec{v}$ is the electron velocity, related to its energy $\epsilon$ through 
$\vec{v}={\nabla}_{\vec{p}} \,\epsilon$; 
%$\vec{F}=e\vec{E}+e \vec{v} \times \vec{B}$ is the force on the electron due to the electric $\vec{E}$ and magnetic $\vec{B}$ fields (note that $e$ includes the sign), 
$\vec{F}=e\vec{E}$ is the force on the electron due to the electric $\vec{E}$ (note that $e$ includes the sign), and $C(f)$ is the linearized collision operator acting on the distribution function. 
For a general collision operator, $C(f)$ writes in terms of the scattering matrix $\mathbf{\Omega}$ as $C(f)= -\Sigma_{\vec{k}'}\Omega_{_{\vec{k}{\vec{k}'}}} n_{\vec{k}'}$, with $n_{\vec{k}}= f_{\vec{k}}-f_{\vec{k}}^{0}$ and $f^{0} \equiv f_{\vec{k}}^{0}$ the Fermi-Dirac distribution.

%The FDF procedure to obtain the constitutive equation and microscopic weights follows prescribed steps. Details can be found in Appendix A.

%\begin{enumerate}[i)]
%\item 
In the present isothermal case, the reduced description is constructed by
retaining the current density,
$\mathbf j=\int e \vec{v} f d\lambda$, with
$d\lambda=d^3k/(4 \pi^3)$, 
together with its first spatial and temporal
derivatives. The nonequilibrium distribution is written as
\begin{equation}
	\label{ansatz_f}
   f=f^{0}+\vec{\beta}_{\vec{k}}\cdot \mathbf j+\vec{\gamma}_{\vec{k}} \cdot \frac{\partial \mathbf j}{\partial t}+\overset{=}{G}_{\vec{k}} : {\nabla}\mathbf j,
\end{equation}
where $\vec{\beta}_{\vec{k}}$, $\vec{\gamma}_{\vec{k}}$ and
$\overset{=}{G}_{\vec k}$ are microscopic coefficient functions associated with
the retained variables.
For simplicity, the subscript $\vec{k}$ will be omitted throughout, while we
restrict the discussion to isotropic materials. In this case one has
$\overset{=}{G}\equiv \overset{=}{G}_{1}+\overset{=}{G}_{2}$, with
$\overset{=}{G}_{1}\equiv g_{1}\hat{k} \hat{k}$ and
$\overset{=}{G}_{2}\equiv g_{2}\mathbb{I}$, where $\hat{k}$ denotes the unit
vector along $\vec{k}$, $\mathbb{I}$ is the identity tensor, and $g_{1}$ and
$g_{2}$ are scalar functions of $\vec{k}$; the symbol ``:'' denotes double
tensor contraction.

The definition of the current density immediately leads to the following
consistency conditions:
\begin{equation}
\label{consistency_beta}
\int e \vec{v} \vec{\beta} d\lambda = \mathbb{I},
\end{equation}
\begin{equation}
\label{consistency_gamma}
\int  \vec{v} \vec{\gamma} d\lambda = 0.
\end{equation}

The macroscopic equations are obtained by projecting the BTE onto appropriate
microscopic quantities. Projection onto the electron charge $e$, together with
particle conservation, yields the charge conservation equation
\begin{equation}
    		e\frac{\partial n}{\partial t}+{\nabla}\cdot \mathbf j=0,
            \label{charge_conservation}
\end{equation}
where $n$ is the electron number density. Projecting onto the microscopic electric current gives an equation for $\mathbf j$:
\begin{equation}
\label{flux_transport}
	\frac{\partial \mathbf j}{ \partial t}+{\nabla}\cdot \overset{=}{\Pi} + \int e \vec{v} \, \vec{F} \cdot  {\nabla}_{\vec{p}} f=\int e  \vec{v} \, C(f)d\lambda,
\end{equation}
with $\overset{=}{\Pi}=\int e \vec{v}\vec{v} f$ the flux of current density.
All integrals extend over the first Brillouin zone.

Equation~(\ref{ansatz_f}) closes the projected equations by expressing the
nonequilibrium distribution in terms of the retained macroscopic variables.
Substituting Eq.~(\ref{ansatz_f}) into Eq.~(\ref{flux_transport}) and
neglecting nonlinear terms yields an equation for the current density:
\begin{equation}
\label{GKE}
\tau\frac{\partial \mathbf j}{\partial t}+\mathbf j=\sigma \vec{E'} +\ell^2\left(\nabla^2\mathbf j+\alpha {\nabla}({\nabla}\cdot \mathbf j)\right),
\end{equation}
where $\vec{E'}=\vec{E}-\nabla \mu/e$ is the total electric field and $\mu$ the
electron chemical potential, $\sigma$ is the electric conductivity, $\tau$ the
current-density relaxation time, $\ell$ the nonlocal length, and $\alpha$ a
nonlocal coefficient. All of them are intrinsic, size-independent
coefficients determined by the microscopic coefficient functions through the following relations:
%We note that Eq.~\eqref{GKE} has the form of the Guyer--Krumhansl equation for phonons. 
%The microscopic expressions for the transport coefficients give:
\begin{subequations}
\begin{align}
\sigma &= \frac{e \int v_x^2 \, \partial_{\epsilon} f_\lambda^{0}d\lambda}{\int v_x C(\beta_{x}) d\lambda}
\label{sigma}
\\
\tau &= \frac{e \int v_x C(\gamma_{x})d\lambda-1}{e \int v_x C(\beta_{x}) d\lambda}
\label{tau}
\\
	\ell^2 &= \frac{\int v_x v_{y} G_{xy} d \lambda}{\int v_x C(\beta_{x}) d\lambda}
	\label{ell}
%\\
%\alpha &= 2+\frac{\int v_{x}^2 \, g_{2} d %\lambda}{\int v_x v_{y} G_{xy} d \lambda%}
%\label{alpha}
\\ 
\alpha &= \frac{\int v_x^2 G_{xx} d\lambda} {\int v_x v_{y} G_{xy} d \lambda} -1.
\label{alpha}
\end{align}
\end{subequations}

\noindent These coefficients can be expressed in other ways. In this form they can directly be applied to three-dimensional (3D) and two-dimensional (2D) systems. Equation~\eqref{GKE} exhibits a characteristic structure: the first term 
$\sigma \vec{E'}$ represents the local conductivity response, while the second 
terms, proportional to second derivatives of $\mathbf j$, capture the 
leading-order nonlocal corrections arising from spatial variations of the 
current. This is the constitutive structure of the
Guyer--Krumhansl equation  originally introduced for nonlocal heat transport
in dielectric crystals \cite{Guyer1966a}. Closely related hydrodynamic equations for charge transport in metals were 
pioneered by Gurzhi \cite{Gurzhi1968} and have been extensively studied in 
recent years in the context of graphene and clean 
metals \cite{Levitov2016,Torre2015}. 
Here, however, Eq.~\eqref{GKE} follows directly from the FDF closure without assuming a hydrodynamic regime. Whether its nonlocal term represents conventional kinetic corrections or collective momentum transport depends on the collision operator. This distinction becomes explicit below for the Callaway model.

%In order to obtain the corresponding equations for the microscopic weight functions, 
%The combination of Eqs. \eqref{BTE}, \eqref{ansatz_f} and transport equations \eqref{charge_conservation}, \eqref{GKE}, neglecting higher-order derivatives, allows to write the Boltzmann equation just in terms of the retained variables. Identification of their coefficients provides the equations for the weight functions (see Appendix A and \cite{Sendra2021}). For general collision operators one has:
By combining Eqs. \eqref{BTE}, \eqref{ansatz_f}, and the transport equations \eqref{charge_conservation}, \eqref{GKE}, we obtain the equations determining the weight functions. The full derivation is presented in Appendix~\ref{app:isothermal_derivation}.

%the constitutive expansion \eqref{ansatz_f} is substituted into the full Boltzmann equation. Then, the transport equations \eqref{charge_conservation}-\eqref{GKE} obtained in the second step are used to write the Boltzmann equation in terms of the fluxes and their derivatives, and the terms involving second- and higher-order spatial and temporal derivatives are neglected, so that only the retained fluxes and their first derivatives survive, consistently with the constitutive expansion itself. Finally, identifying the coefficients of these independent variables provides the equations determining the weight functions. 
%The weight functions $\vec{\beta}$, $\vec{\gamma}$, and $\overset{=}{G}$ are determined from the Boltzmann equation through a systematic procedure detailed in Appendix A.
\begin{subequations}
    \begin{align}
	C(\vec{\beta})&=\frac{e}{\sigma} \vec{v} \, \partial_{\epsilon} f^{0}
    \label{beta} \\
	C(\vec{\gamma})&=\vec{\beta} + \frac{e\tau}{\sigma} \vec{v} \,\partial_{\epsilon} f^{0}
	%- \tau C(\vec{\beta}_\lambda)
    \label{gamma} \\
C(\overset{=}{G}_{1})&= \vec{v} \,\vec{\beta}
\label{g1} \\
	C(g_{2})&=\frac{1}{e} \left(\frac{\partial \mu}{\partial n}\right)_{T} \! \partial_{\epsilon} f^{0}.
    \label{g2}
    \end{align}
\end{subequations}
These equations are essentially the same as found for phonons \cite{Sendra2021} by changing
$1/\kappa \rightarrow e/\sigma$, $\partial_{T} f^{0}\rightarrow -\partial_{\epsilon} f^{0}$ and $1/C_v \rightarrow 1/e \left(\frac{\partial \mu}{\partial n}\right)_{T}$, with $\kappa$ the thermal conductivity and $C_v$ the heat capacity per unit volume. \textcolor{black}{The equilibrium distribution correspondingly changes from the Planck to the Fermi--Dirac distribution.}

Let us note that the nonequilibrium distribution \eqref{ansatz_f} with weight functions obeying Eqs. \eqref{beta}-\eqref{g2} satisfies the conservation of the number of electrons for general collision operators, i.e. $\int C( f_\lambda) d\lambda=0$. 
%This allows to apply the formalism to approximate collision operators. 
For simple collision models this conservation may be automatic, whereas for more complex approximate operators it need not be trivial. In those cases, the ansatz allows the use of approximate collision operators while preserving particle conservation. %{\it In phonon transport, in contrast, particle density is not conserved. There, the FDF guaranties energy conservation in collisions in approximate models like RTA and Callaway models with mode-dependent relaxation times, which is at odds with the direct application of these approaches, where energy conservation requires additional constraints \cite{Vermeersch2015,Hua2019}.}

We next specialize these general results to the Callaway model, which distinguishes momentum-relaxing from momentum-conserving collisions and therefore allows the same constitutive framework to be followed continuously from the kinetic resistive limit to the collective regime. 

The general derivation above makes no assumption regarding the degeneracy of the electron gas. In contrast, evaluating explicit transport coefficients for specific collision models, as done in the remainder of this section, relies on the sharp Fermi surface (degenerate) approximation, appropriate for metals or highly doped semiconductors \cite{Ziman1963}.

\subsection{Callaway model}\label{sec:Callaway}

The Callaway model was formulated originally in the context of phonon heat transport in dielectrics \cite{Callaway1959}. Its central physical ingredient is the distinction between momentum-relaxing (Resistive) and momentum-conserving (Normal) collisions. In phonon systems, the regime in which Normal collisions dominate over Resistive ones provides the classical setting for hydrodynamic heat transport \cite{Guyer1966a,Guyer1966b}. A closely analogous hierarchy occurs for electrons. Scattering with impurities, phonons, and other momentum sinks relaxes the electronic momentum, whereas normal electron--electron collisions redistribute momentum among the electrons while conserving the total momentum. When the latter become sufficiently frequent compared with momentum-relaxing processes, collective electronic flow can emerge \cite{Bandurin2016}. Callaway's model and closely related variants have therefore also been used for electron transport \cite{Estrada24,Estrada25,Estrada25b,Lian2024}.

For the present purpose, an important feature of the Callaway model is that it is not restricted to the hydrodynamic regime. By varying the relative strength of the two collision mechanisms, it connects continuously the kinetic resistive limit, in which electron--electron collisions are negligible, with the collective regime in which momentum-conserving collisions dominate. It therefore provides a convenient setting in which to examine whether the same FDF constitutive equation describes both regimes and the crossover between them.

Callaway model considers that resistive collisions drive the distribution function to local equilibrium, and that momentum conserving collisions lead the distribution to one which conserves momentum density $\vec{g}=\int \hbar \vec{k} f d\lambda$. The latter can be obtained from the Boltzmann equation or from entropy maximization, and provides the so-called displaced distribution
%\begin{equation}
%    f^{disp}=\frac{1}{exp(\epsilon-\mu-\hbar \vec{k}\!\cdot \! \vec{u})/kT+1} 
%\end{equation}
\begin{equation}
f^{disp}=\frac{1}{e^{(\epsilon-\mu-\hbar \vec{k}\cdot \vec{u})/k_B T}+1},
\end{equation}
with $k_B$ the Boltzmann constant, $T$ the temperature, and $\vec{u}$ a parameter with velocity units related to the momentum density.  Close to equilibrium, one can keep to first order
\begin{equation}
f^{disp}=f^0 - \hbar \, \vec{k} \!\cdot \! \vec{u}\, \partial_{\epsilon} f^{0}.
\end{equation}

The collision operator in Callaway model is
\begin{equation}
\label{Callaway}
    C(f)=- \frac{f-f^0}{\tau_R}-  \frac{f-f^{disp}}{\tau_{ee}},
\end{equation}
where $\tau_R$ is the average scattering time for momentum-relaxing processes, such as electron--phonon and impurity scattering, which contribute to electrical resistance, while $\tau_{ee}$ is the average
collision time for momentum-conserving electron--electron scattering. 
In the degenerate regime ($k_BT\ll\epsilon_F$, with $\epsilon_F$ the Fermi
energy), only states within a narrow energy window around the Fermi level
contribute to transport. If the collision times vary weakly over this
window, they can be approximated by their values at the Fermi level, so
that $\tau_R$ and $\tau_{ee}$ become independent of $\vec{k}$.

For a general isotropic dispersion relation, the coefficients of
Eq.~\eqref{GKE} can then be evaluated analytically, yielding (see
Appendix~\ref{app:callaway_singleflux})
\begin{equation}
  \tau=\tau_R, \hspace{5mm} \ell^2=\frac{v_F^2 \tau_R \tau_C}{d+2}, \hspace{5mm} \alpha=\frac{d-2}{d},
\end{equation}
where $v_F$ is the Fermi velocity, $\tau_C^{-1}=\tau_R^{-1}+\tau_{ee}^{-1}$ is the total collision rate, and $d$ is the system dimension. The resulting constitutive equation predicts nonlocal
corrections also in the regime dominated by momentum-relaxing collisions,
showing that nonlocality is not restricted to the collective hydrodynamic
regime, consistent with previous results for two-dimensional systems \cite{Estrada24,Estrada25,Baker2024}. Here the same constitutive structure follows for three-dimensional degenerate electron systems. The dimensional dependence is also noteworthy: in two dimensions $\alpha=0$, so the term proportional to $\nabla(\nabla\cdot\mathbf j)$ vanishes, whereas in one dimension $1+\alpha=0$, so the two second-gradient contributions cancel identically for a one-dimensional flow. 

Introducing
\begin{equation}
 r=\frac{\tau_C}{\tau_R},
 \label{r_Callaway}
\end{equation}
the crossover between the kinetic and collective regimes becomes
transparent. When electron--electron collisions are negligible,
$\tau_{ee}\gg\tau_R$, one has $\tau_C\simeq\tau_R$ and
$r\rightarrow1$. 
The Callaway operator then reduces to the usual kinetic
resistive limit, and the nonlocal length becomes
$\ell=v_F\tau_R/\sqrt{d+2}$. Thus, nonlocal corrections persist even
when momentum-relaxing collisions dominate. Its magnitude can be estimated directly from tabulated Fermi velocities
and resistive relaxation times \cite{Gall2016}. At room temperature,
representative values are $\ell\simeq24$, $18$, and $17$~nm for Ag, Cu,
and Au, respectively, while Al gives a smaller value,
$\ell\simeq8$~nm because of its shorter resistive relaxation time.
These estimates show that nonlocal length scales of several to a few
tens of nanometers are expected even in ordinary metals outside the
strongly collective regime.
In the opposite limit, $\tau_{ee}\ll\tau_R$, one has $r\ll1$: momentum
is redistributed many times by electron--electron collisions before being
relaxed resistively, and the system enters the collective regime. The same
nonlocal length then reduces to the usual Gurzhi length,
$\ell_G=\sqrt{\ell_{ee}\ell_{\rm imp}/(d+2)}$, with
$\ell_{ee}=v_F\tau_{ee}$ and $\ell_{\rm imp}=v_F\tau_R$
\cite{Gurzhi1968,Bandurin2016,Guo2017_Pnas,Estrada24,Narozhny2022}.

The corresponding nonequilibrium distribution is
\begin{equation}
\label{f_Callaway}
    f=f^{0} - \frac{e\tau_R}{\sigma} \, \partial_{\epsilon} f^{0} \, \vec{v} 
    \cdot \mathbf j+ 
    \frac{e \tau_R \tau_C}{\sigma} \partial_{\epsilon} f^{0} (\vec{v} \vec{v} - \frac{v^2}{d} \mathbb{I}) : {\nabla}\mathbf j,
\end{equation}
which, for a horizontal flow such as Couette or Poiseuille, can be cast as 
\begin{equation}
\label{Couette_Callaway}
    f\simeq f^{0}(\vec{k}-\frac{\tau_R e}{\sigma \hbar} j_x \hat{k}_x + 
   \frac{\tau_R \tau_C e}{\sigma \hbar}  v_y \frac{\partial j_x}{\partial y} \hat{k}_x).
\end{equation}
The explicit form of the distribution function provides a simple microscopic picture of the two roles played by the retained variables. The current $j_x$ shifts the Fermi surface, whereas the transverse gradient $\partial j_x/\partial y$ produces an additional anisotropic distortion. The latter represents the redistribution of electronic momentum between neighboring layers that underlies the viscous-like spatial response.

The Callaway model therefore makes explicit that hydrodynamic transport emerges as one limit of the FDF kinetic constitutive equation, rather than being introduced as a separate macroscopic description. In the following section, we apply the same procedure to electronic heat transport, derive the corresponding nonlocal constitutive equation, and compare its predictions with thermal-transport measurements in metallic nanostructures.
%We next test the nonlocal constitutive description against electronic heat-transport measurements in metallic nanostructures.

\section{Experimental validation in metallic thermal transport}
\label{sec:thermal_validation}

Size-dependent electronic heat transport in copper films and in copper
and nickel inverse opals at room temperature can be described by the FDF heat-current equation
within the relaxation-time approximation (RTA), without invoking
momentum-conserving electron--electron collisions or assuming a
hydrodynamic electron fluid. We first derive the electronic heat-current equation and later compare to experiments.

\subsection{Heat-current equation}

\begin{figure*}
    \centering
\IfFileExists{Fig1_JT.png}{\includegraphics[width=0.99\linewidth]{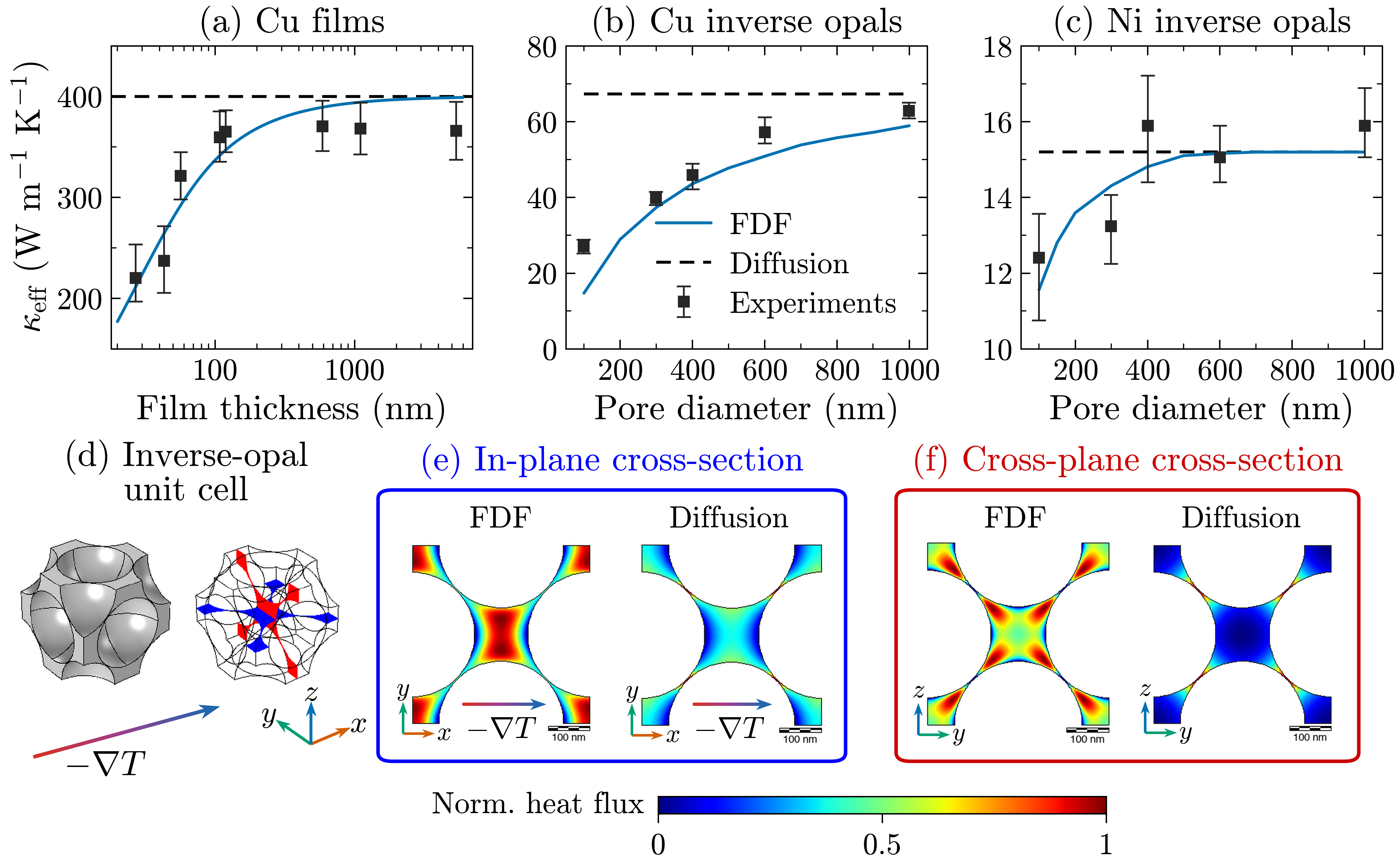}}{} % F1.jpg not present in the supplied source package
    \caption{Apparent thermal conductivity and heat flux profiles in metallic nanostructures at room temperature. The FDF predictions are compared with experimental data \cite{barako2016quasi,islam2024}; the standard Fourier prediction is also shown as a reference. (a) Apparent thermal conductivity of copper films as a function of thickness. (b,c) Apparent thermal conductivity of copper and nickel inverse opals, respectively, as a function of pore diameter for the fully packed FCC geometry illustrated in (d). (e,f) Heat-flux profiles, normalized by their maximum value, across the blue (in-plane) and red (cross-plane) cross-sections indicated in (d) for a Cu inverse opal with a pore size of 300~nm. The FDF predicts heat-flux boundary layers near the free surfaces and a reduction of the energy current along curved streamlines, accounting for the decrease of the apparent conductivity as the nanostructure size is reduced.}
    \label{F1}
\end{figure*}

For purely thermal transport, the retained flux is the electronic heat
current
\begin{equation}
 \mathbf q_e=\int(\epsilon-\mu)\vec v\,f\,d\lambda.
 \label{heat_flux_single}
\end{equation}
The corresponding FDF representation of the distribution is thus
\begin{equation}
 f=f^0+\vec{\beta}_q\!\cdot\mathbf q_e
 +\vec{\gamma}_q\!\cdot\partial_t\mathbf q_e
 +\overset{=}{G}_q:\nabla\mathbf q_e,
 \label{thermal_fdf_ansatz}
\end{equation}
and in the relaxation-time approximation the collision term is
\begin{equation}
 C(f)=-\frac{f-f^0}{\tau_R}.
 \label{thermal_rta_collision}
\end{equation}
In the degenerate isotropic limit, the corresponding thermal weights give
\begin{equation}
 \ell_q^2=\frac{v_F^2\tau_R^2}{d+2}.
 \label{ellq_RTA}
\end{equation}
The evaluation of the thermal weight functions leading to
Eq.~\eqref{ellq_RTA} is given in
Appendix~\ref{app:callaway_singleflux}. \textcolor{black}{The RTA is thus the kinetic limit of the Callaway model discussed in Sec.~\ref{sec:Callaway}.}

In general, electronic energy is not conserved separately because electrons can exchange energy with the phonon subsystem. In stationary linear response, conservation of the total heat current gives
\begin{equation}
 \nabla\cdot(\mathbf q_e+\mathbf q_{\rm ph})=0,
\label{energy_conservation_thermal_v2}
\end{equation}
so that $\nabla\cdot\mathbf q_e=-\nabla\cdot\mathbf q_{\rm ph}$. If the phonon heat current is locally described by Fourier's law, $\mathbf q_{\rm ph}=-\kappa_{\rm ph}\nabla T$, the longitudinal second-gradient contribution scales as $\nabla(\nabla\cdot\mathbf q_e)\sim\kappa_{\rm ph}\Delta T/L^3$, whereas $\nabla^2\mathbf q_e\sim\kappa_e\Delta T/L^3$. Their relative magnitude is therefore of order $\kappa_{\rm ph}/\kappa_e$. Since heat transport in the metallic systems considered below is predominantly electronic at room temperature, with $\kappa_{\rm ph}\ll\kappa_e$, we neglect this longitudinal correction and in steady state get
\begin{equation}
 \nabla\cdot\mathbf q_e =0
\label{energy_conservation_thermal}
\end{equation}
\begin{equation}
 \textcolor{black}{\mathbf q_e=-\kappa_e\nabla T+\ell_q^2\nabla^2\mathbf q_e.}
 \label{GK_thermal_FDF}
\end{equation}
Hereafter, since only the electronic heat current is retained, we denote
$\mathbf q_e$ simply by $\mathbf q$ and $\kappa_e$ by $\kappa$.
Equation~\eqref{GK_thermal_FDF} has the Guyer--Krumhansl form, but its use
here does not imply that the electrons are in the Gurzhi regime. To our
knowledge, Eq.~\eqref{GK_thermal_FDF} is the first Guyer--Krumhansl-type
constitutive equation for electronic heat transport derived directly from
the electron Boltzmann equation. Previous electronic Guyer--Krumhansl
equations were formulated within extended thermodynamic
frameworks~\cite{Jou2012,Machrafi2014,Sellitto2016}.

In the RTA, the nonlocal scale is fixed by the microscopic collision time, so that boundary layers may modify the measured heat current even in the absence of momentum-conserving electron--electron collisions. 
The boundary conditions are derived from the kinetic description of electron scattering at the surface, rather than imposed phenomenologically. For the gray Callaway model considered here, this derivation was given by Guo and Wang~\cite{Guo2018}. The corresponding derivation for coupled charge and heat transport is presented in Appendix~\ref{app:boundary_conditions}, with the single-flux condition recovered as a particular case. At an
insulating free surface we impose
\begin{equation}
    \mathbf q\cdot\mathbf n=0,
    \label{n}
\end{equation}
where $\mathbf n$ is the outward unit vector normal to the surface, while the tangential component satisfies a Maxwell-type slip condition 
\begin{equation}
 q_t=-\ell_{\rm slip}\,\partial_n q_t.
 \label{thermal_slip_bc}
\end{equation}
For diffuse boundary scattering, the Maxwell model gives $\ell_{\rm slip}\simeq1.19\,\ell_q$ \cite{Guo2018}, so that the boundary condition introduces no independent adjustable length scale. 
%The general kinetic boundary conditions and their derivation are discussed in Appendix~\ref{app:boundary_conditions}.

Equations~\eqref{energy_conservation_thermal}--\eqref{thermal_slip_bc}
define the boundary-value problem used below for the Cu films and the Cu and Ni inverse opals. We solve this problem below using standard finite-element methods \cite{Beardo2019}.

\subsection{Copper films and metallic inverse opals}

We first consider the apparent thermal conductivity of copper films as a
function of thickness. When the film thickness becomes comparable to the
electronic mean free path, the conductivity decreases from its bulk value.
Within the FDF, this size dependence arises from the reduction
of the heat flux near the free surfaces. 

For copper at room temperature, we use the intrinsic parameters
$\kappa_{\rm Cu}=400~{\rm W\,m^{-1}K^{-1}}$ and
$\ell_q=17.4~{\rm nm}$ \cite{barako2016quasi}
The resulting FDF prediction reproduces the observed decrease in
conductivity with decreasing film thickness \cite{islam2024}, as shown in
Fig.~\ref{F1}(a), whereas the local Fourier description contains no
corresponding size dependence.

Inverse opals provide a substantially more stringent geometrical test.
The metallic matrix contains an FCC array of spherical voids, forcing the
heat current to bend around curved internal surfaces. The problem is
therefore genuinely multidimensional and cannot be reduced to a
one-dimensional suppression factor.

For each periodic unit cell we solve
Eqs.~\eqref{energy_conservation_thermal} and
\eqref{GK_thermal_FDF}, imposing a temperature difference $\Delta T$
between two opposite terminals and the diffuse-wall conditions \eqref{n}--\eqref{thermal_slip_bc} at the
pore surfaces. The apparent conductivity is obtained from the integrated
terminal heat current,
\begin{equation}
 \kappa_{\rm eff}
 =\frac{L}{S\Delta T}
 \int_{\Gamma}\mathbf q\cdot d\vec S,
 \label{kappa_eff_opal}
\end{equation}
where $S$ is the terminal area and $L$ the unit-cell length.

For copper we use the same intrinsic parameters as in the film
calculation. For nickel we use
$\kappa_{\rm Ni}=87.9~{\rm W\,m^{-1}K^{-1}}$ and
$\ell_q=2.7~{\rm nm}$ \cite{barako2016quasi}. The calculated apparent conductivity decreases
systematically as the pore size is reduced, in agreement with the
experimental data of Barako \textit{et al.}~\cite{barako2016quasi}, as
shown in Fig.~\ref{F1}(b,c). Within the FDF, this reduction is associated
with the spatial redistribution of the heat flux illustrated in
Fig.~\ref{F1}(d--f), including boundary layers near the free surfaces and
curved heat-flow paths through the metallic network.

The measured size dependence has previously been described using ballistic transport models \cite{islam2024,barako2016quasi}. The present results show that the same experimental trends can also be reproduced by the FDF using a nonlocal constitutive equation and the corresponding boundary conditions. This provides a complementary reduced description of the observed size effects, formulated directly in terms of the heat flux.

The films and inverse opals provide complementary benchmarks. The former
probe size-dependent reduction in a simple transverse geometry, whereas
the latter test the same constitutive law in a complex three-dimensional
boundary-value problem. This comparison also illustrates a practical
advantage of the FDF: once the kinetic information has been reduced to
bulk transport coefficients and boundary conditions, nonlocal transport
can be treated as a macroscopic boundary-value problem and solved with
standard numerical methods even in complex geometries, where a direct
solution of the Boltzmann equation would be considerably more demanding.

The RTA also yields a direct electrical prediction from this thermal benchmark. To leading order in the degenerate limit, the charge and heat currents have the same nonlocal length and the same normalized diffuse-wall boundary condition. Consequently, for any given geometry,
\begin{equation}
\frac{\sigma_{\rm eff}}{\sigma_{\rm bulk}}
=
\frac{\kappa_{\rm eff}}{\kappa_{\rm bulk}}
+O[(T/T_F)^2],
\label{eq:WF_finite_size_RTA}
\end{equation}
so that the finite-size correction preserves the Wiedemann--Franz relation at this order. Electrical measurements on the same structures would therefore provide an independent, parameter-free experimental test of the nonlocal constitutive description.

\section{Coupled thermoelectric transport}
\label{sec:thermoelectric}

The results discussed so far concern charge and heat transport separately.
In thermoelectric transport, however, the two currents are coupled: an
electric field and a temperature gradient generally drive both
$\mathbf j$ and $\mathbf q$. Once spatial nonlocality is retained, this
coupling can also involve spatial variations of the two currents, so that
the separate constitutive equations derived above are no longer sufficient.
This motivates treating $\mathbf j$ and $\mathbf q$ simultaneously as the
nonequilibrium variables of the FDF closure.

\subsection{Coupled constitutive formulation}
\label{sec:coupled_fdf}
Accordingly, the stationary FDF closure is extended to retain both the electric and electronic heat currents:
\begin{equation}
f=f^0+
\vec{\beta}_{j}\cdot\mathbf j+
\vec{\beta}_{q}\cdot\mathbf q+
\overset{=}{G}_j:\nabla\mathbf j+
\overset{=}{G}_q:\nabla\mathbf q .
\label{eq:ansatz_thermoelectric}
\end{equation}
Following the same projection procedure as in Appendix~\ref{app:isothermal_derivation}, we obtain coupled constitutive equations for the charge and heat currents. For stationary transport, charge conservation gives $\nabla\cdot\mathbf j=0$ exactly. The electronic heat current, however, need not be separately conserved [see Eq.~\eqref{energy_conservation_thermal_v2}]. As discussed in Sec.~IV, when $\kappa_{\rm ph}\ll\kappa_e$, the longitudinal contribution $\nabla(\nabla\cdot\mathbf q_e)$ can be neglected to leading order under the conditions considered here. We therefore retain only the Laplacian contribution in the heat-current sector and denote the electronic heat current simply by $\mathbf q$. The resulting coupled constitutive equations are
\begin{equation}
\mathbf j=
L_{11}\mathbf E
+L_{12}\left(-\frac{\nabla T}{T}\right)
+M_{11}\nabla^2\mathbf j
+M_{12}\nabla^2\mathbf q
\label{eq:j_coupled}
\end{equation}
and
\begin{equation}
\mathbf q=
L_{21}\mathbf E
+L_{22}\left(-\frac{\nabla T}{T}\right)
+M_{21}\nabla^2\mathbf j
+M_{22}\nabla^2\mathbf q .
\label{eq:q_coupled}
\end{equation}
%The coupled projection is formulated in compact matrix form in ec.~\ref{sec:formal_solution}.
Equations~\eqref{eq:j_coupled} and \eqref{eq:q_coupled} constitute a central result of this work: charge and heat currents obey a coupled nonlocal constitutive law in which spatial variations of either current can affect both responses.

The coefficients $L_{ij}$ describe the conventional local thermoelectric response. Thus $L_{11}$ and $L_{22}$ determine the direct electrical and thermal responses, while $L_{12}$ and $L_{21}$ describe the local thermoelectric coupling between charge and heat transport. The matrix $M$ introduces a qualitatively different coupling associated with spatial variations of the currents themselves. The diagonal coefficients $M_{11}$ and $M_{22}$ describe the nonlocal electrical and thermal responses already encountered in the corresponding single-flux problems. More importantly, the off-diagonal coefficients $M_{12}$ and $M_{21}$ couple the spatial structure of the two currents: a nonuniform heat current contributes to the electric current through $M_{12}\nabla^2\mathbf q$, whereas a nonuniform electric current contributes to the heat current through $M_{21}\nabla^2\mathbf j$.

This distinction is physically important. In local thermoelectric transport, charge and heat currents are coupled only through the thermodynamic forces. In the FDF, they are additionally coupled through their spatial profiles. \textcolor{black}{Thus confinement and boundary-induced current profiles can produce size-dependent corrections to the electrical, thermal, and thermoelectric response.}
Having established the physical content of the coupled equations, it is convenient to introduce the flux and force vectors
\[
\mathbf J=
\begin{pmatrix}
\mathbf j\\
\mathbf q
\end{pmatrix},
\qquad
\mathbf X=
\begin{pmatrix}
\mathbf E\\
-\nabla T/T
\end{pmatrix}.
\]

Denoting by $L$ and $M$ the $2\times2$ matrices with elements $L_{ij}$ and $M_{ij}$, respectively, the coupled constitutive equations take the compact form
\begin{equation}
\mathbf J=L\mathbf X+M\nabla^2\mathbf J.
\label{eq:coupled_FDF}
\end{equation}

The construction leading to Eq.~\eqref{eq:coupled_FDF} also determines the
kinetic weights entering the nonequilibrium distribution and the transport
coefficients $L$ and $M$. Although these quantities can be written directly
as momentum integrals, doing so becomes cumbersome when charge and heat
transport are treated simultaneously. A kinetic-space notation provides a
more compact formulation of the FDF and leads naturally to operator
expressions for $L$ and $M$. We introduce this formulation in the next
subsection and use it throughout the remainder of the paper.

\subsection{Kinetic-space formulation}
\label{sec:kinetic_matrix}

We write the distribution as
\begin{equation}
f=f^0+\left(-\partial_\epsilon f^0\right)\phi .
\label{eq:deltaf_phi}
\end{equation}
The function $\phi(\vec p)$ and the microscopic currents can be regarded as vectors in kinetic space. For each spatial component $i$, we define the charge- and heat-current vectors \cite{Lucas2018}
\begin{subequations}\label{eq:current_vectors}
\begin{align}
|J_i\rangle
&=\int d\lambda\, e\,v_i(\vec p)|\mathbf p\rangle,
\label{eq:current_vector_J}\\
|Q_i\rangle
&=\int d\lambda\,(\epsilon-\mu)v_i(\vec p)|\mathbf p\rangle .
\label{eq:current_vector_Q}
\end{align}
\end{subequations}
The kinetic space is endowed with the equilibrium-weighted inner product
\begin{equation}
\langle\psi|\phi\rangle
=
\int \psi(\vec p)\phi(\vec p)
\left(-\partial_\epsilon f^0\right)d\lambda .
\label{eq:kinetic_inner_product}
\end{equation}
With this definition, the physical current integrals are simply
\begin{equation}
j_i=\langle J_i|\phi\rangle,
\qquad
q_i=\langle Q_i|\phi\rangle .
\label{eq:physical_currents_projection}
\end{equation}

It is then convenient to collect the two microscopic current vectors as
\begin{equation}
A_i=
\begin{pmatrix}
|J_i\rangle & |Q_i\rangle
\end{pmatrix},
\qquad
A_i^T=
\begin{pmatrix}
\langle J_i|\\[1mm]
\langle Q_i|
\end{pmatrix}.
\label{eq:A_definition}
\end{equation}
Thus
\begin{equation}
\begin{pmatrix}
j_i\\
q_i
\end{pmatrix}
=A_i^T|\phi\rangle .
\label{eq:J_projection}
\end{equation}
In what follows, the spatial index is left implicit when no ambiguity arises, and the same notation $A$ is used in the compact kinetic equations.

The reduced perturbation $\phi$ has the same flux-dependent structure as the perturbation in Eq.~\eqref{eq:ansatz_thermoelectric}, with weights obtained by dividing those defined earlier by the equilibrium factor $-\partial_\epsilon f^0$. Collecting the local weights in $\beta=(\vec{\beta}_j,\vec{\beta}_q)$ and denoting by $G_a$ the corresponding first-gradient weights associated with the spatial derivative $\partial_a$, the requirement that this representation reproduce the physical currents gives the consistency conditions
\begin{equation}
A^T\beta=I,\qquad A^TG_a=0.
\label{eq:consistency_matrix_FDF}
\end{equation}
The first relation ensures that the local part of the distribution carries exactly the prescribed charge and heat currents, while the second ensures that the gradient correction does not redefine them.

Under the equilibrium-weighted product~\eqref{eq:kinetic_inner_product}, the linearized collision operator is self-adjoint. Denoting the corresponding adjoint by $\dagger$,
\begin{equation}
C^\dagger=C.
\label{C_selfadjoint_FDF}
\end{equation}
Thus, the Hermiticity of $C$ refers throughout to this equilibrium-weighted kinetic product; no redefinition of the physical quantities entering the charge and heat fluxes is implied.

\subsection{Kinetic derivation of the coupled thermoelectric equations}
\label{sec:formal_solution}

In this notation, the FDF derivation takes a particularly compact form and proceeds as follows. We first project the Boltzmann equation
to obtain the coupled constitutive equations in terms
of the kinetic weights. We then use both the Boltzmann
equation and the resulting constitutive equations to
determine these weights, allowing us to express the
transport matrices directly in terms of microscopic
kinetic quantities.

\textcolor{black}{With the electric and thermal driving terms collected in $-A\mathbf X$, the stationary linearized Boltzmann equation can be written as}
\begin{equation}
C\phi=-A\mathbf X+V_a\partial_a\phi ,
\label{eq:BTE_compact}
\end{equation}
where $V_a$ denotes multiplication by the velocity component $v_a$. Within the FDF, the nonequilibrium function is written in terms of the physical currents and their first spatial derivatives,
\begin{equation}
\phi=\beta\mathbf J+G_a\partial_a\mathbf J .
\label{eq:FDF_distribution}
\end{equation}
The weights $\beta$ describe the local distortion of the distribution associated with the charge and heat currents, whereas $G_a$ describes the additional distortion generated when these currents vary in space.

\textcolor{black}{Projecting Eq.~\eqref{eq:BTE_compact} with $A^T$, defining}
\begin{equation}
B=A^TA ,
\label{eq:B_definition}
\end{equation}
and substituting Eq.~\eqref{eq:FDF_distribution} gives an equation directly for $\mathbf J$. For the isotropic systems considered here, $A$ and $\beta$ are odd under momentum inversion, while $G_a$ is even. Since the collision operator preserves parity, the terms containing first spatial derivatives vanish upon projection,
\begin{equation}
A^TCG_a=0,
\qquad
A^TV_a\beta=0.
\label{eq:first_gradient_parity}
\end{equation}
The projected Boltzmann equation therefore reduces to
\begin{equation}
A^TC\beta\,\mathbf J
=
-B\mathbf X
+A^TV_aG_b\,\partial_a\partial_b\mathbf J .
\label{eq:BTE_projected_reduced}
\end{equation}

Comparison with the constitutive equation
\begin{equation}
\mathbf J=L\mathbf X+M_{ab}\partial_a\partial_b\mathbf J
\label{eq:FDF_tensor}
\end{equation}
then gives
\begin{equation}
L=(-A^TC\beta)^{-1}B ,
\label{eq:L_beta}
\end{equation}
and
\begin{equation}
M_{ab}=-LB^{-1}A^TV_aG_b .
\label{eq:M_G}
\end{equation}
The local response is therefore associated with the part of the distribution proportional to $\mathbf J$, whereas the nonlocal response originates from the change in the distribution produced by spatial variations of the physical currents.

For isotropic systems, the second-gradient operator in Eq.~\eqref{eq:FDF_tensor} consists of a Laplacian contribution and a longitudinal term proportional to $\nabla(\nabla\cdot\mathbf J)$. The latter vanishes for the stationary charge current and is neglected for the electronic heat current under the assumption of negligible phononic contribution to heat transport, as discussed in Sec.~\ref{sec:coupled_fdf}. Consequently, Eq.~\eqref{eq:FDF_tensor} reduces to the Laplacian form of Eqs.~\eqref{eq:j_coupled} and \eqref{eq:q_coupled}, with $M$ denoting the matrix multiplying $\nabla^2\mathbf J$.

It remains to determine the kinetic weights $\beta$ and $G_a$.
We substitute Eq.~\eqref{eq:FDF_distribution} for $\phi$
into the Boltzmann equation and eliminate $\mathbf X$
using the constitutive equation \eqref{eq:FDF_tensor}.
Retaining terms up to first order in spatial derivatives
for the determination of the weights, and matching the
terms proportional to $\mathbf J$ and
$\partial_a\mathbf J$, gives
\begin{equation}
C\beta=-AL^{-1},
\qquad
CG_a=V_a\beta .
\label{eq:weights_equations}
\end{equation}
These relations generalize Eqs.~\eqref{beta}--\eqref{g2} to the stationary coupled problem.
We define
\begin{equation}
R=-C^{-1},
\qquad R^\dagger=R .
\label{eq:R_definition}
\end{equation}
Since the collision invariants belong to the null space of $C$, the inverse $C^{-1}$ is understood to act on the subspace orthogonal to these invariants. The kinetic weights are therefore
\begin{equation}
\beta=RAL^{-1},
\qquad
G_a=-RV_aRA L^{-1}.
\label{eq:beta_G_solution}
\end{equation}

The consistency condition $A^T\beta=I$ then gives
\begin{equation}
L=A^TRA ,
\label{eq:L_general}
\end{equation}
the standard kinetic-theory expression for the local transport matrix
\cite{Ziman1963,LucasHartnoll2018}. Finally, substituting $G_b$ from Eq.~\eqref{eq:beta_G_solution} into Eq.~\eqref{eq:M_G} gives
\begin{equation}
M_{ab}
=
LB^{-1}A^TV_aRV_bRA L^{-1},
\qquad
B=A^TA .
\label{eq:M_general}
\end{equation}
Thus, the FDF extends the standard expression for the local matrix in Eq.~\eqref{eq:L_general} to the nonlocal matrix M.

\subsection{Callaway collision model}\label{sec:Callaway_coupled}
\label{sec:coupled_callaway}

We now specialize the general expressions above to the Callaway collision model introduced in Sec.~III, Eq.~\eqref{Callaway}. In the kinetic-space notation, the momentum-conservation condition determines the displacement velocity from the projection of the nonequilibrium distribution onto the momentum vectors
\begin{equation}
|P_i\rangle=\int d\lambda\,p_i\,|\mathbf p\rangle,
\qquad
u_i=\frac{\langle P_i|\phi\rangle}{\langle P_i|P_i\rangle}.
\label{eq:callaway_u_projection}
\end{equation}
The displaced part of the distribution is therefore the projection of $|\phi\rangle$ onto the momentum sector. Introducing the projector operators
\begin{equation}
\mathcal P=\sum_i
\frac{|P_i\rangle\langle P_i|}{\langle P_i|P_i\rangle},
\qquad
\mathcal Q=I-\mathcal P,
\label{eq:callaway_projectors}
\end{equation}
the Callaway collision operator of Sec.~\ref{sec:Callaway} can be written as
\begin{equation}
C=-\frac{1}{\tau_R}\mathcal P
  -\frac{1}{\tau_C}\mathcal Q,
\qquad
\frac{1}{\tau_C}=\frac{1}{\tau_R}+\frac{1}{\tau_{ee}}.
\label{eq:callaway_C_projector}
\end{equation}
Thus the momentum sector is relaxed only by resistive collisions, whereas the orthogonal sector is relaxed by both scattering mechanisms. Since $\mathcal P$ and $\mathcal Q$ are orthogonal projectors, the inverse entering the FDF coefficients follows directly,
\begin{equation}
R=-C^{-1}=\tau_R\mathcal P+\tau_C\mathcal Q .
\label{eq:callaway_R}
\end{equation}
Substitution of Eq.~\eqref{eq:callaway_R} into
Eqs.~\eqref{eq:L_general} and \eqref{eq:M_general} provides the coupled
local and nonlocal transport matrices for the Callaway model. 
As in Sec.~\ref{sec:Callaway}, we take the relaxation times to be energy independent, an
approximation particularly well suited to the degenerate regime, where
transport is restricted to a narrow energy window around the Fermi
surface. In terms of $r=\tau_C/\tau_R$ the resistive regime corresponds
to $r\simeq1$, while $r\ll1$ produces a strong separation between the
slowly relaxing momentum sector and the remaining kinetic degrees of
freedom.

The structure of Eq.~\eqref{eq:callaway_R} also provides a simple
connection with previous results for local interacting transport. Using
$\mathcal P+\mathcal Q=I$, $R$ can be written as an interpolation between
its resistive and collective limits,
\begin{equation}
R=rR_{\rm RTA}+(1-r)R_{\rm col},
\label{eq:R_Callaway_interpolation}
\end{equation}
where $R_{\rm RTA}=\tau_R I$ and
$R_{\rm col}=\tau_R\mathcal P$ are the resistive and collective
limits, respectively. Since $L=A^T R A$, the same interpolation
follows immediately for the local transport coefficients,
\begin{equation}
L_{ij}=rL_{ij}^{\rm RTA}+(1-r)L_{ij}^{\rm col}.
\label{eq:L_Callaway_interpolation}
\end{equation}
Thus, the weighted interpolation between the two limits follows directly
at the operator level, without requiring the individual transport
coefficients to be evaluated. The same structure for the thermal and
thermoelectric conductivities was obtained by Lucas and Das Sarma from a
more detailed low-temperature kinetic calculation
\cite{Lucas2018}.

For a large Fermi surface, $L_{11}^{\rm col}\simeq\tau_R\langle J_x|J_x\rangle=L_{11}^{\rm RTA}\simeq L_{11}$, so that, to leading order in the degenerate expansion, the electrical conductivity is not modified by electron--electron collisions. The thermal and thermoelectric coefficients behave differently. Since the relevant projection $\langle P_x|Q_x\rangle$ is of order $(k_BT/\epsilon_F)^2$, the thermal conductivity is reduced to $\kappa\simeq(\tau_C/\tau_R)\kappa_{\rm RTA}$, and hence $\mathcal L\equiv\kappa/(\sigma T)\simeq L_0\tau_C/\tau_R$, with $L_0=(\pi^2/3)(k_B/e)^2$ the Sommerfeld value of the Lorenz number. Thus momentum-conserving electron--electron scattering suppresses the Lorenz ratio below its RTA value. The thermoelectric coefficients are likewise modified, leading also to deviations from the Mott relation. 

Violations of the Wiedemann--Franz and Mott relations in the hydrodynamic regime have previously been obtained using more detailed kinetic treatments~\cite{Lucas2018}. Hydrodynamic thermoelectric transport in confined graphene devices has also been studied using macroscopic fluid descriptions~\cite{LiCorbino2022,LiHallbar2022}. Recent mesoscopic heat-transport experiments have likewise reported substantial deviations from the Wiedemann--Franz law in the hydrodynamic regime~\cite{Pusep2026}.

\subsection{Degenerate isotropic bands}
\label{sec:degenerate_isotropic}

We now consider the specific case of degenerate isotropic 
bands and
evaluate the nonlocal transport matrix $M$. The Sommerfeld evaluation of
the kinetic averages entering both $L$ and $M$ is given in
Appendix~\ref{app:M_isotropic}. For the energy-independent Callaway model
considered here, the resulting coupled constitutive response satisfies
Onsager reciprocity to the Sommerfeld order retained below, as shown in
Appendix~\ref{app:bulk_reciprocity}.

To separate the universal Fermi-surface structure from the detailed band dispersion $\epsilon=\epsilon(p)$, we introduce the spectral functions
\begin{equation}
a(\epsilon)=\frac{D(\epsilon)v^2(\epsilon)}{d},\qquad
b(\epsilon)=\frac{p(\epsilon)}{v(\epsilon)},\qquad
c(\epsilon)=\frac{v^2(\epsilon)}{d+2}.
\label{eq:abc_general}
\end{equation}
We also define
\begin{equation}
\ell_C^2\equiv\tau_C^2c_F
=\frac{v_F^2\tau_C^2}{d+2},
\label{eq:ellC_general}
\end{equation}
where the subscript $F$ denotes evaluation at the Fermi energy. Evaluating the general kinetic expression, Eq.~\eqref{eq:M_general}, through the retained Sommerfeld order gives
\begin{equation}
M=\ell_C^2
\begin{pmatrix}
r^{-1}&m_{12}\\
m_{21}&1
\end{pmatrix},
\label{eq:M_general_isotropic}
\end{equation}
where
\begin{subequations}\label{eq:mij_general}
\begin{align}
m_{12}&=e\left[
r^{-1}\frac{c'}{c}
+(r^{-1}-1)\frac{b'}{b}
\right]_{F},\\
m_{21}&=\frac{\pi^2(k_BT)^2}{3e}
\left[
\frac{c'}{c}
+(r^{-1}-1)
\left(\frac{a'}{a}+\frac{b'}{b}\right)
\right]_{F}.
\end{align}
\end{subequations}
The derivation of Eqs.~\eqref{eq:M_general_isotropic} and \eqref{eq:mij_general}, including the required kinetic projections and Sommerfeld expansion, is given in Appendix~\ref{app:M_isotropic}.

A first result follows directly from the diagonal elements. At leading
degenerate order, for an arbitrary isotropic dispersion,
\begin{equation}
\begin{split}
M_{11}&=\frac{\ell_C^2}{r}
=\frac{v_F^2\tau_C\tau_R}{d+2},\\M_{22}&=\ell_C^2
=\frac{v_F^2\tau_C^2}{d+2}.
\end{split}
\label{eq:M_diagonal_scales}
\end{equation}
These two coefficients have a simple kinetic interpretation. In the degenerate limit, the charge-current mode coincides, to leading Sommerfeld order, with the momentum mode for an isotropic band. Momentum-conserving collisions therefore do not relax this mode, which has a lifetime $\tau_R$, although both momentum-conserving and resistive collisions contribute to its spatial redistribution on the shorter time $\tau_C$. The heat-current sector, by contrast, is controlled at this order by the complementary kinetic modes, for which both relaxation and spatial redistribution occur on the time $\tau_C$. The corresponding nonlocal lengths can thus be viewed as diffusion lengths, \textcolor{black}{$\ell^2\sim D\tau_{\rm life}$}, with an effective diffusion coefficient $D\sim v_F^2\tau_C$. This gives the mixed scale $v_F^2\tau_C\tau_R$ in $M_{11}$, whereas $M_{22}\propto v_F^2\tau_C^2$. These are the same kinetic scales encountered in the separate charge- and heat-current problems of Secs.~III and IV. At this order, both diagonal coefficients are independent of the detailed band dispersion, whereas the band dependence enters the off-diagonal coefficients $M_{12}$ and $M_{21}$.

To make this band dependence explicit, consider the power-law dispersion
\begin{equation}
\epsilon=A p^z.
\label{eq:power_law_dispersion}
\end{equation}
Substitution into Eq.~\eqref{eq:mij_general} gives
\begin{subequations}\label{eq:mij_power_law}
\label{pz}
\begin{align}
m_{12}^{(z)}&=
\frac{e}{z\epsilon_F}
\left[z(r^{-1}+1)-2\right],\\
m_{21}^{(z)}&=
\frac{\pi^2(k_BT)^2}{3ez\epsilon_F}
\left[d r^{-1}+2z-d-2\right].
\end{align}
\end{subequations}
Their consequences for linear and quadratic bands will be discussed in
Sec.~\ref{sec:strip_BTE_benchmark}, where the different microscopic origins of the finite-size thermoelectric response become explicit.

The relation between the two-current closure and descriptions based on a
single momentum mode, such as the reduced closure of
Ref.~\cite{Coulter2025}, is particularly transparent for a parabolic
dispersion. In this case, the charge current is directly proportional to
momentum, so that retaining the momentum mode amounts to retaining
$\mathbf j$ as the single constitutive flux. In the present notation, this
corresponds to omitting the
$M_{12}$ and $M_{22}$ terms. Importantly, the $M_{22}\nabla^2\mathbf q$ term provides the bulk nonlocal heat-current contribution entering the description of the Cu and Ni thermal experiments discussed above. Its omission therefore removes this thermal finite-size mechanism from the momentum-only closure. This difference becomes particularly relevant in finite geometries, where the constitutive equations must be supplemented by boundary conditions. A boundary condition on the retained momentum mode constrains that mode directly and, for a parabolic band, also the proportional charge current, whereas the boundary response associated with the eliminated heat-current sector is not retained explicitly. By retaining both physical currents, the FDF enables the bulk constitutive equations and the kinetic boundary conditions, derived in Appendix~\ref{app:boundary_conditions}, to be expressed in terms of the same fields, providing a unified boundary-value formulation for finite-size thermal and thermoelectric transport.

\section{Physical consequences of the coupled constitutive equations}
\label{sec:confined_transport}

\textcolor{black}{The coupled constitutive equations derived in Sec.~V imply more generally that confinement modifies the complete transport response, rather than only individual thermoelectric coefficients. Once supplemented by the kinetic boundary conditions, they define a geometry-dependent relation between the section-averaged currents and the applied forces, $\overline{\mathbf J}=L_{\rm eff}[\mathcal G]\mathbf X$, where $\mathcal G$ denotes the sample geometry. Thus the electrical, thermal, and thermoelectric coefficients all become effective properties of the boundary-value problem. We first examine how this nonlocality appears directly in the coupled constitutive equations, focusing on the Seebeck and Peltier configurations, before formulating the complete geometry-dependent transport matrix.}

\subsection{Isothermal transport: charge flow and Peltier response}
\label{sec:isothermal_transport}

We first consider uniform temperature, $\nabla T=0$. The coupled constitutive equations, Eqs.~\eqref{eq:j_coupled} and \eqref{eq:q_coupled}, become
\begin{subequations}\label{eq:isothermal_coupled}
\begin{align}
\mathbf j&=L_{11}\mathbf E+M_{11}\nabla^2\mathbf j+M_{12}\nabla^2\mathbf q,\label{eq:j_isothermal}\\
\mathbf q&=L_{21}\mathbf E+M_{21}\nabla^2\mathbf j+M_{22}\nabla^2\mathbf q.\label{eq:q_isothermal}
\end{align}
\end{subequations}
To leading local order under isothermal electrical driving, $\mathbf q\simeq L_{21}\mathbf E$. Since $L_{21}=O(T^2)$, while $M_{12}=O(T^0)$ in the Sommerfeld counting, the feedback term $M_{12}\nabla^2\mathbf q$ in Eq.~\eqref{eq:j_isothermal} contributes to the charge-current equation only at $O(T^2)$. The charge-current equation therefore reduces, to leading order in the degenerate expansion, to
\begin{equation}
\mathbf j=L_{11}\mathbf E+\ell_j^2\nabla^2\mathbf j,
\qquad
\ell_j^2\equiv M_{11}=\frac{v_F^2\tau_C\tau_R}{d+2}.
\label{eq:stokes_ohm_FDF}
\end{equation}
In the collective limit $r\ll1$, Eq.~\eqref{eq:stokes_ohm_FDF} takes the familiar Stokes--Ohm form of viscous electron flow. More generally, however, the same flux-gradient equation follows from the FDF without assuming the collective regime. Away from that limit, its nonlocal term represents kinetic spatial corrections rather than hydrodynamic viscosity.

The same isothermal problem shows how a nonuniform charge-current profile enters the Peltier response. Eliminating the electric field between Eqs.~\eqref{eq:j_isothermal} and \eqref{eq:q_isothermal} gives
\begin{equation}
\mathbf q=\Pi_0\mathbf j
+\left(M_{21}-\Pi_0M_{11}\right)\nabla^2\mathbf j
+\left(M_{22}-\Pi_0M_{12}\right)\nabla^2\mathbf q,
\label{eq:peltier_exact}
\end{equation}
where
\begin{equation}
\Pi_0=\frac{L_{21}}{L_{11}}=T S_0.
\end{equation}
{$\Pi_0$ is the local Peltier coefficient.}
Thus the heat current is not determined {solely} by $\Pi_0\mathbf j$: both the curvature of the electrical current and the independent spatial dynamics of the heat current contribute. This is the Peltier manifestation of retaining $\mathbf j$ and $\mathbf q$ as independent constitutive variables.

To isolate the direct curvature contribution, we use the local relation $\mathbf q\simeq\Pi_0\mathbf j$ as a leading-order estimate only inside the last term of Eq.~\eqref{eq:peltier_exact}. For the power-law dispersion $\epsilon=Ap^z$, this gives, to the retained Sommerfeld order,
\begin{equation}
\mathbf q\simeq\Pi_0\left(\mathbf j+\ell_\Pi^2\nabla^2\mathbf j\right),
\label{eq:peltier_reduced}
\end{equation}
with
\begin{equation}
\ell_\Pi^2=\ell_C^2
\frac{z(1+r)-2r}{d+r(z-2)}.
\label{eq:ellPi_general}
\end{equation}
For $z\geq 1$, the direct curvature contribution satisfies $\ell_\Pi^2\geq 0$. In the resistive limit (RTA, $r=1$), it vanishes for a linear dispersion ($z=1$) but remains finite for a quadratic dispersion ($z=2$). {Equation~\eqref{eq:peltier_reduced} then shows that, for the current curvature considered here, this nonlocal contribution reduces the Peltier heat.} Equation~\eqref{eq:ellPi_general} identifies the microscopic scale associated with this contribution, but does not by itself determine the finite-size Peltier response, which follows from the full coupled boundary problem in Sec.~\ref{sec:finite_size_reciprocity}.

\subsection{Open-circuit transport: heat flow and Seebeck response}
\label{sec:open_circuit_transport}

We now consider a conductor subject to a temperature gradient under open-circuit conditions. The electrical constraint is on the section-averaged current,
\begin{equation}
\overline j=0,
\label{eq:open_circuit_average}
\end{equation}
not on $j(\mathbf r)$ pointwise. Imposing $\mathbf j(\mathbf r)=0$ pointwise is in general incompatible with the electrostatic constraint $\nabla\times\mathbf E=0$. Local electrical counterflows are therefore required, with their contributions canceling upon section averaging. The general compatibility condition and the exceptional cases permitting pointwise $\mathbf j=0$ are discussed in Appendix~\ref{app:counterflow_closure}.

To determine the heat-current profile, we eliminate the electric field from the two coupled constitutive equations without setting the local charge current to zero. This gives
\begin{equation}
\begin{split}
\mathbf q={}&\Pi_0\mathbf j-\kappa_{\rm oc}\nabla T
+\left(M_{21}-\Pi_0M_{11}\right)\nabla^2\mathbf j\\
&+\left(M_{22}-\Pi_0M_{12}\right)\nabla^2\mathbf q,
\end{split}
\label{eq:heat_twoflux_open}
\end{equation}
with
\begin{equation}
\kappa_{\rm oc}=\frac{1}{T}
\left(L_{22}-\frac{L_{21}L_{12}}{L_{11}}\right).
\label{eq:kappa_oc_main}
\end{equation}
{\color{black}
Equation~\eqref{eq:heat_twoflux_open} makes explicit the feedback of
electrical counterflows on heat transport.  Its smallness is particularly
transparent in the resistive limit, $r=1$, relevant to the Cu/Ni
thermal benchmark.  A temperature gradient generates an electrical response
of scale $\sigma S\nabla T$; its feedback into the heat current through the
Peltier coefficient $\Pi_0=TS$ therefore changes the thermal response by
$\delta\kappa_{\rm TE}\sim\sigma S^2T$.  Using the Wiedemann--Franz
relation, $\kappa_e\simeq L_0\sigma T$, gives
$\delta\kappa_{\rm TE}/\kappa_e\sim S^2/L_0=O[(T/T_F)^2]$, beyond the
leading degenerate thermal response.  Appendix~\ref{app:thermal_benchmark}
shows explicitly that the same leading-order result follows from the full
coupled equation for $0<r\leq1$.  Thus, \textcolor{black}{for arbitrary $r$,}
\begin{equation}
\mathbf q=-\kappa_{\rm oc}\nabla T+\ell_C^2\nabla^2\mathbf q,
\qquad
\ell_C^2=\frac{v_F^2\tau_C^2}{d+2}.
\label{eq:heat_GK_main}
\end{equation}

{We now turn to the electrical response.} To determine how this heat-current profile enters the Seebeck response, we isolate the electric field from the first coupled constitutive equation:}

\begin{equation}
\mathbf E=S_0\nabla T
+\frac{1}{L_{11}}
\left(
\mathbf j-M_{11}\nabla^2\mathbf j-M_{12}\nabla^2\mathbf q
\right),
\label{eq:seebeck_general}
\end{equation}
where
\begin{equation}
S_0=\frac{L_{12}}{T L_{11}}
\label{eq:S0_def}
\end{equation}
is the bulk Seebeck coefficient.

\begin{figure}[t]
\centering
\includegraphics[width=0.98\columnwidth]{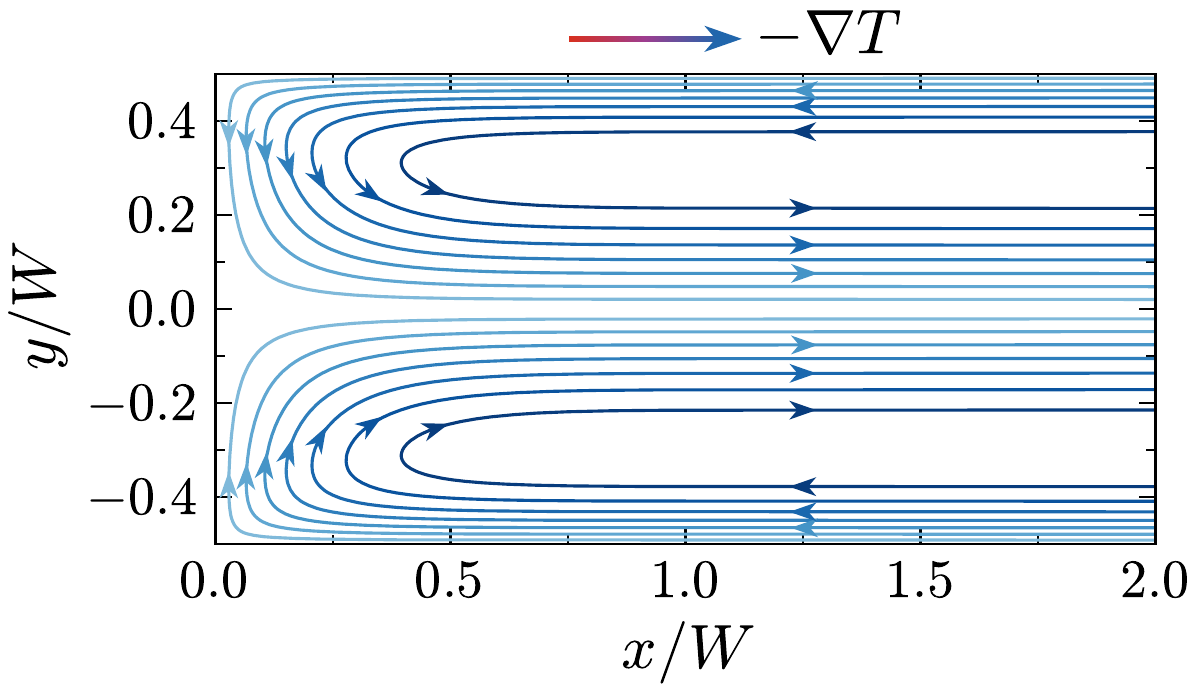}
\caption{Closure of electrical counterflows \(\mathbf j(\mathbf r)\) under a thermal gradient in an open-circuit strip. Electrical-current streamlines obtained from the finite-geometry coupled FDF for a two-dimensional (\(d=2\)) parabolic band (\(z=2\)), with \(r=\tau_C/\tau_R=1\) and \(W=v_F\tau_C\). For an electron-like Seebeck response (\(S_0<0\)), with the hot end on the left, the electrical current flows toward the cold end at the strip center and returns toward the hot end near the walls. These oppositely directed currents close through transverse currents near the electrically insulating ends. Only the hot end of the strip is shown; the closure near the cold end is equivalent.}
\label{fig:counterflow_loops_app}
\end{figure}

The heat-current profile determined by Eq.~\eqref{eq:heat_GK_main} enters the electrical equation through the $M_{12}\nabla^2\mathbf q$ term. Equation~\eqref{eq:seebeck_general} is therefore the nonlocal counterpart of the classical Seebeck relation: the electric field required to maintain open circuit depends on the spatial structure of both currents. The $M_{12}\nabla^2\mathbf q$ term provides a direct heat-profile contribution to the Seebeck response, while the electrical-current profile $\mathbf j(\mathbf r)$, including the $M_{11}\nabla^2\mathbf j$ contribution, describes the counterflows generated by the same boundary problem. Figure~\ref{fig:counterflow_loops_app} illustrates the resulting electrical-current streamlines for a finite strip, obtained by solving the coupled FDF equations together with charge conservation and electrically insulating boundary conditions. Details of the finite-geometry calculation are given in Appendix~\ref{app:counterflow_closure}.

For the power-law dispersion $\epsilon=A p^z$, the direct heat-profile contribution is controlled by
\begin{equation}
-\frac{M_{12}}{L_{11}}
=-\frac{\ell_C^2}{e\tau_C a_F\epsilon_F}
\frac{z(1+r)-2r}{z},
\label{eq:seebeck_cross_general}
\end{equation}
As for the Peltier contribution discussed above, this term vanishes for a linear dispersion in the resistive limit $r=1$, whereas for a quadratic dispersion it remains finite at this order. Since $S_0=L_{12}/(T L_{11})$, its relative strength is governed by $-T M_{12}/L_{12}$. For the power-law dispersion, $M_{12}/L_{12}\propto\ell_\Pi^2$, so the same microscopic combination appears in the direct Seebeck and Peltier curvature corrections. The measurable finite-size Seebeck coefficient, however, is determined by the full coupled boundary problem and cannot in general be identified with the $M_{12}$ term alone. The full relation between the effective Seebeck and Peltier coefficients follows instead from reciprocity of the coupled boundary problem, as shown below.

\subsection{Finite-size thermoelectric response and reciprocity}
\label{sec:finite_size_reciprocity}

For a specified geometry and kinetic boundary condition, the coupled FDF equations define a linear boundary-value problem. Its section-averaged response can therefore be written in the same thermodynamic form as the bulk relation,
\begin{equation}
\overline{\mathbf J}=L_{\rm eff}[\mathcal G]\,\mathbf X,
\label{eq:Leff_geometry}
\end{equation}
where $\mathcal G$ denotes the geometry and its boundary conditions. The matrix $L_{\rm eff}$ contains the complete effect of the nonlocal current profiles, including the mutual coupling of $\mathbf j$ and $\mathbf q$ and the local electrical counterflows allowed under open circuit. It is this matrix, rather than an individual element of $M$, that determines the measurable finite-size thermoelectric coefficients.
Thus $L_{\rm eff}$ provides a unified description of the electrical, thermal, and thermoelectric response of the confined system. The thermal examples discussed above show that its finite-size consequences can be substantial, with changes of tens of percent as transverse dimensions approach the nonlocal scale. We now turn to the nonlocal Seebeck and Peltier effects. Under open circuit, $\overline j=0$, Eq.~\eqref{eq:Leff_geometry} gives
\begin{equation}
S_{\rm eff}[\mathcal G]
=\frac{[L_{\rm eff}]_{12}}
{T[L_{\rm eff}]_{11}},
\label{eq:Seff_geometry}
\end{equation}
whereas under isothermal electrical driving
\begin{equation}
\Pi_{\rm eff}[\mathcal G]
=\frac{[L_{\rm eff}]_{21}}
{[L_{\rm eff}]_{11}}.
\label{eq:Pieff_geometry}
\end{equation}
The bulk coefficients $S_0$ and $\Pi_0$ are recovered when the characteristic transverse dimensions become large compared with the kinetic lengths.

The bulk FDF response satisfies Onsager reciprocity to the Sommerfeld order retained here, and the diffuse kinetic boundary condition preserves the corresponding reciprocal structure, as shown in Appendices~\ref{app:bulk_reciprocity} and \ref{app:boundary_conditions}. Hence
\begin{equation}
[L_{\rm eff}]_{12}=[L_{\rm eff}]_{21},
\end{equation}
and the confined coefficients obey the Kelvin relation
\begin{equation}
\Pi_{\rm eff}[\mathcal G]=T S_{\rm eff}[\mathcal G]
\label{eq:kelvin_confined}
\end{equation}
{Since the Kelvin relation holds both in the bulk and for the confined response, the relative nonlocal corrections satisfy}
\begin{equation}
\frac{\delta\Pi}{\Pi_0}
=\frac{\delta S}{S_0}.
\label{eq:relative_kelvin_confined}
\end{equation}
This result is stronger than the approximate profile arguments above: it applies to the full coupled solution and does not require replacing $\mathbf q$ by $\Pi_0\mathbf j$ in the Peltier problem or neglecting the electrical counterflows in the Seebeck problem.

{The physical consequence is simple. Boundaries reshape the two coupled currents and thereby generate geometry-dependent effective Seebeck and Peltier coefficients. The Kelvin relation requires their relative finite-size corrections to be identical, even though the two experiments excite very different current profiles.}

\subsection{Differential experimental signatures}
\label{sec:experimental_signatures}

The geometry dependence of $S_{\rm eff}$ and $\Pi_{\rm eff}$ suggests differential configurations in which the common bulk response cancels. The basic idea is to use two branches made of the same material but with very different transverse dimensions: one branch is macroscopic, while the other approaches the kinetic length scales. This avoids comparing different materials and turns the finite-size correction itself into the measured signal.

\begin{figure}[t]
    \centering
    \includegraphics[width=0.75
\columnwidth]{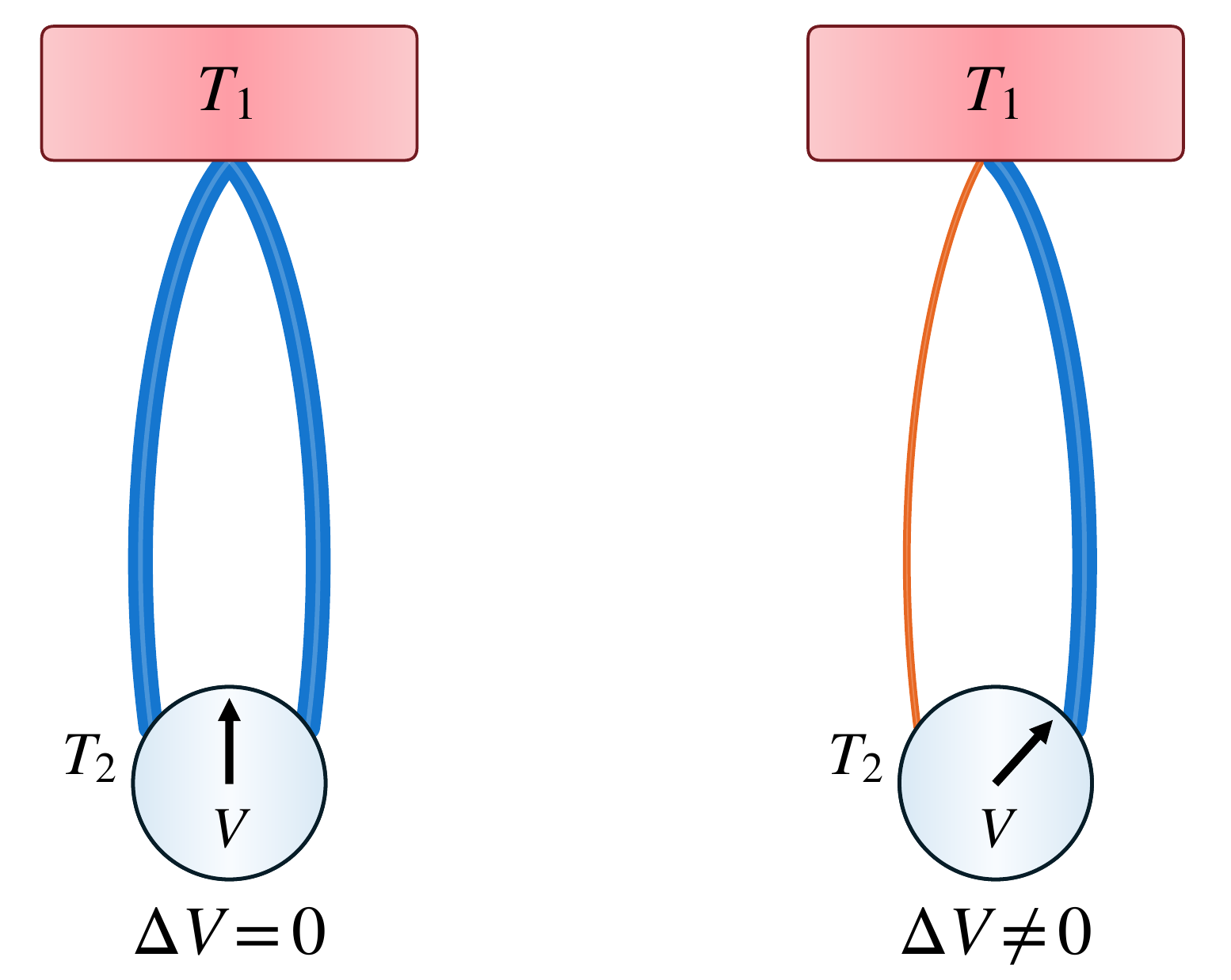}
    \caption{Differential Seebeck effect in wires made of the same metal. (Left) When both branches are thick, their effective Seebeck coefficients are identical and the measured electromotive force vanishes. (Right) When one branch has a transverse size comparable to the kinetic lengths while the other remains thick, their finite-size Seebeck coefficients differ and a nonzero differential voltage is generated. The thick branch provides an internal bulk reference, so that the common bulk thermopower cancels.}
    \label{fig:seebeck_differential}
\end{figure}

\paragraph{Differential Seebeck effect.}
{Consider two branches of the same material connecting the same hot and cold reservoirs and connected to a voltmeter, as shown in Fig.~\ref{fig:seebeck_differential}.} If both branches are macroscopic and otherwise equivalent, their Seebeck voltages are identical and the measured electromotive force vanishes. If one branch remains macroscopic while the other is confined, the common bulk contribution still cancels, but the confined branch has a different effective thermopower. {The differential signal is therefore}
\begin{equation}
\mathcal E_{\rm diff}
=\int_{T_c}^{T_h}
\left[S_{\rm eff}^{\rm thin}(T)-S_{\rm eff}^{\rm thick}(T)\right]dT,
\label{eq:seebeck_differential}
\end{equation}
and for a sufficiently thick reference branch $S_{\rm eff}^{\rm thick}\simeq S_0$. {Several asymmetric units could in principle be connected in series to increase the voltage signal.}

Equation~\eqref{eq:seebeck_general} shows microscopically why such a difference can arise: confinement generates spatial structure in both $\mathbf q$ and the open-circuit counterflow $\mathbf j$, so the electric field needed to enforce zero net current differs from its bulk value. The term $M_{12}\nabla^2\mathbf q$ identifies the direct heat-profile channel, but Eq.~\eqref{eq:seebeck_differential} involves the complete coupled response and {does not assign the measured signal to that term alone.}

The microscopic origin of the finite-size response depends on the band dispersion. For a linear dispersion, the direct heat-profile coefficient $M_{12}$ vanishes in the resistive limit and a finite-size thermopower appears when {the momentum-conserving and resistive relaxation times are separated.} For a parabolic dispersion, a finite-size response is already present in the resistive limit. The full finite-width calculations below show that these qualitative distinctions survive after all coupled terms are included.

\begin{figure}[t]
    \centering
    \includegraphics[width=0.9\columnwidth]{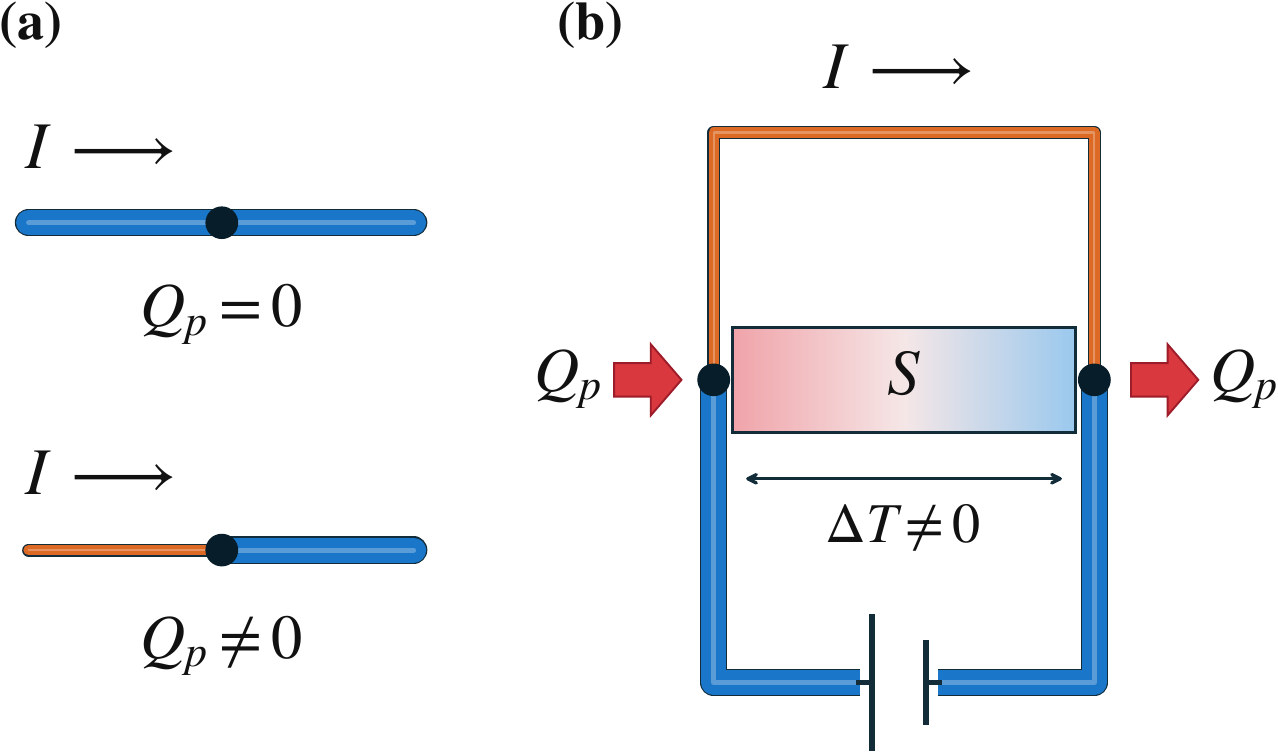}
    \caption{Differential Peltier effect in wires made of the same metal. (a, top) When both wires are thick, their Peltier coefficients are identical and no net Peltier heat is exchanged with the environment at the junction, $Q_P=0$. (a, bottom) When one wire has a transverse size comparable to $\ell_C$ while the other remains thick, their finite-width Peltier coefficients differ and a nonzero Peltier heat $Q_P=I(\Pi_{\rm thin}-\Pi_{\rm thick})$ is absorbed or released at the junction. (b) A possible configuration in which the corresponding heating and cooling generate a temperature difference across an electrically insulating sample $S$.}
    \label{fig:peltier_differential}
\end{figure}

\paragraph{Differential Peltier effect.}
{The reciprocal configuration, shown in Fig.~\ref{fig:peltier_differential}, is obtained by driving an electrical current through a junction between two wires of the same material.} Here $Q_P$ denotes the net Peltier heat exchanged with the environment at the junction. If both wires are thick, their Peltier coefficients are identical and $Q_P=I(\Pi_1-\Pi_2)=0$, even though each wire carries a Peltier heat current. If one wire is confined while the other remains macroscopic, their finite-size Peltier coefficients differ and the same-metal junction absorbs or releases
\begin{equation}
Q_P=I\left(\Pi_{\rm eff}^{\rm thin}-\Pi_{\rm eff}^{\rm thick}\right).
\label{eq:peltier_differential}
\end{equation}
By Eq.~\eqref{eq:kelvin_confined}, this is the reciprocal counterpart of the differential Seebeck response, with the same relative finite-size correction for the same geometry. {Figure~\ref{fig:peltier_differential}(b) sketches how the associated heating and cooling can generate a temperature difference across an electrically insulating sample.}

The confined branch need not be a uniform wire or strip. Constrictions, smoothly varying sections, or curved geometries can be used as long as the kinetic boundary condition remains applicable. The film considered next is introduced for a different purpose: it provides a controlled geometry in which the full coupled FDF can be compared directly with the spatially resolved Boltzmann equation.

\subsection{Kinetic benchmark}
\label{sec:strip_BTE_benchmark}

{We now consider a strip of thickness $W$, infinite along $x$, with transverse coordinate $y$, for which the longitudinal FDF reads}
\begin{equation}
\mathbf J(y)=L\mathbf X+M\,\partial_y^2\mathbf J(y),
\label{eq:strip_FDF}
\end{equation}
with the kinetic boundary conditions derived in Appendix~\ref{app:boundary_conditions}. Open circuit is imposed on the integrated current,
\begin{equation}
\overline j=\frac{1}{W}\int_{-W/2}^{W/2}j(y)\,\dd y=0,
\label{eq:average_open_circuit}
\end{equation}
so local electrical counterflows remain possible. The solution of the coupled thermoelectric equations \eqref{eq:j_coupled}--\eqref{eq:q_coupled} and the resulting $S_{\rm eff}(W)$ are given in Appendix~\ref{app:seebeck_benchmark}, together with the spatially resolved Boltzmann calculation used for comparison.

\begin{figure*}[t]
\centering
{\includegraphics[width=1.9\columnwidth]{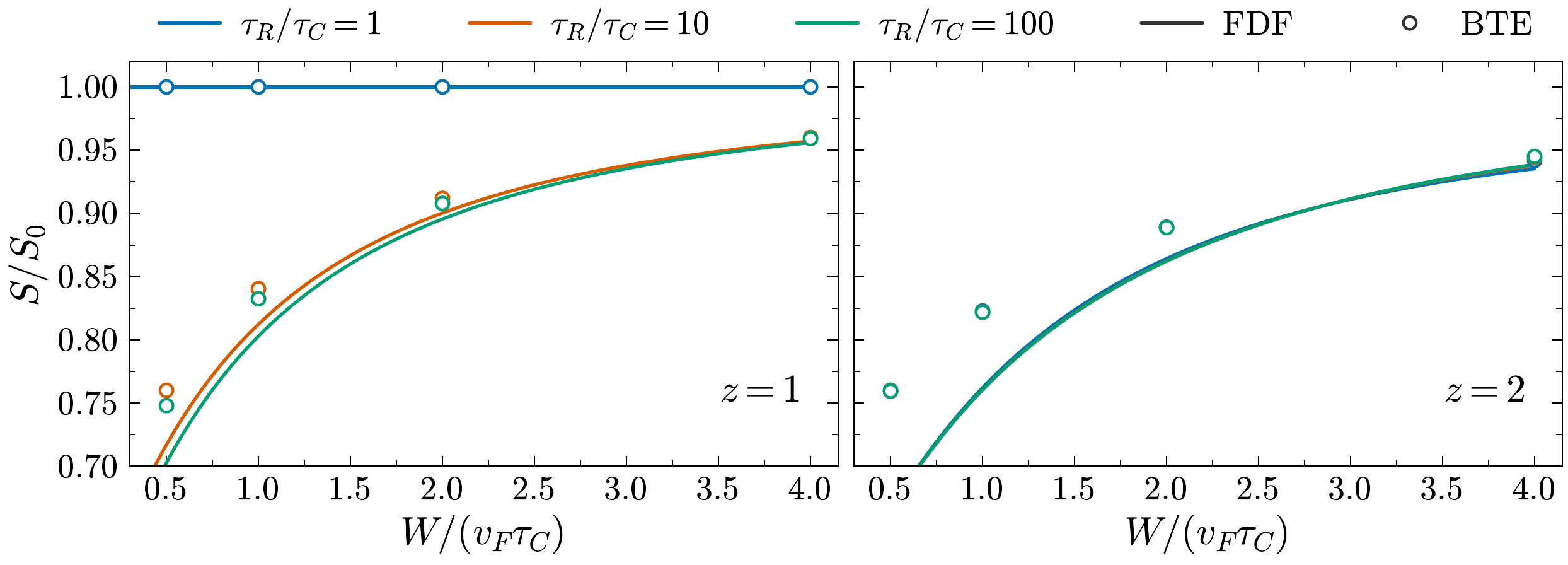}}
\caption{Finite-width Seebeck response used as a kinetic benchmark of the FDF for a two-dimensional strip of thickness $W$. Curves are the complete coupled FDF predictions and symbols are spatially resolved BTE results as the relative strength of momentum-conserving and resistive scattering is varied through $\tau_R/\tau_C=1/r$. Left: linear dispersion ($z=1$). Right: parabolic dispersion ($z=2$); the FDF curves and BTE symbols for the different values of $r$ overlap. The agreement improves systematically with increasing $W/(v_F\tau_C)$. The strip is used as a controlled benchmark of the constitutive closure, not as a privileged experimental geometry.}
\label{fig:widthscan_main}
\end{figure*}

Figure ~\ref{fig:widthscan_main} compares the Seebeck coefficient predicted by the coupled FDF with that obtained from the full spatially resolved BTE in two dimensions, considering two representative dispersions: linear ($z=1$) and parabolic ($z=2$).
{Around the kinetic crossover, $W\gtrsim v_F\tau_C$, the FDF captures the effective thermopower reasonably while the finite-size correction remains visible, and the agreement with the BTE improves progressively for larger widths.}

{For both dispersions, the finite-size effect reduces the Seebeck coefficient relative to its bulk value. Through the confined Kelvin relation of Sec.~VI.C, the same relative reduction applies to the effective Peltier coefficient, consistent with the sign of the direct curvature contribution identified in Sec.~VI.B.}

{The two dispersions, however, display a markedly different dependence on $r$. For a linear dispersion, the reduction increases as momentum-conserving scattering becomes more important and vanishes in the RTA ($r=1$). In contrast, for a parabolic dispersion, both the FDF and BTE results are essentially independent of $r$, and the finite-size reduction survives even at $r=1$.}

{This contrasting dependence on $r$, examined in more detail in Appendix~\ref{app:seebeck_benchmark}, reflects different microscopic origins of the size dependence. For $z=1$, the carrier velocity is independent of energy. In the RTA, the boundary therefore introduces the same energy-independent geometric factor in $L_{11}^{\rm eff}$ and $L_{12}^{\rm eff}$, leaving their ratio, and hence the Seebeck coefficient, unchanged. When the momentum-conserving and resistive relaxation times differ, this common factorization is broken by the collective momentum response, producing a finite-size correction to their ratio. For $z=2$, the carrier velocity depends on energy, so the boundary affects carriers at different energies differently and modifies the Seebeck coefficient already in the RTA.}

{The structure of $M_{12}$ provides a useful microscopic marker of these two mechanisms.} Rewriting the coefficient obtained in Sec.~\ref{sec:coupled_callaway} as
\begin{equation}
m_{12}^{(z)}=
\frac{2e(z-1)}{z\epsilon_F}
+\frac{e}{\epsilon_F}\frac{1-r}{r},
\label{eq:m12_two_origins}
\end{equation}
separates a band-dispersion contribution that survives at $r=1$ from a contribution generated by the separation of $\tau_C$ and $\tau_R$. For $z=1$ only the second term remains; for $z=2$ the first term is already finite in RTA. This mirrors the two kinetic mechanisms identified independently by the full BTE. The agreement in Fig.~\ref{fig:widthscan_main}, however, benchmarks the complete coupled FDF response, including the counterflows and all elements of $M$, rather than the $M_{12}$ channel alone.

\section{Conclusions and Discussion}
\label{sec:conclusions}

We have developed the Flux Derivative Formalism (FDF) as a kinetic
constitutive framework for coupled nonlocal charge and heat transport.
Starting directly from the linearized Boltzmann equation, the FDF uses the
physical transport fluxes as the nonequilibrium variables of the reduced
description and constructs the distribution function from these fluxes and
their leading derivatives. This leads to a compact set of coupled
constitutive equations in which charge and heat currents are treated on
equal footing, retaining both their local response and their mutual
nonlocal coupling. Together with the boundary conditions derived
consistently within the same reduced description, these equations define a
macroscopic boundary-value problem directly in terms of the physical
currents. This reduction makes the framework amenable to complex geometries
using standard macroscopic numerical methods, without requiring the full
phase-space Boltzmann problem to be solved for each geometry. 

Within the
Callaway description considered here, the form of the nonlocal constitutive
equations is preserved as the relative importance of resistive and
momentum-conserving collisions is varied, while the collision physics enters
through the transport coefficients. The same framework therefore provides
a continuous description across the crossover from resistive-dominated
transport to the collective hydrodynamic limit
\cite{Estrada24,Estrada25}.

The coupled equations recover familiar transport descriptions in the appropriate limits while also exposing genuinely nonlocal thermoelectric effects. In the isothermal limit the charge-current equation reduces to the Stokes--Ohm form of viscous electronic transport \cite{Gurzhi1968,Sulpizio2019}, whereas under open-circuit conditions the same framework yields a nonlocal heat equation of the Guyer--Krumhansl type together with a geometry-dependent Seebeck response. Importantly, open circuit constrains the net electrical current rather than the local current, generally producing local electrical counterflows whose contributions cancel upon cross-section averaging.

For degenerate electron systems and the Callaway model, the coupled
constitutive coefficients satisfy Onsager reciprocity to the retained
Sommerfeld order, while the boundary problem preserves the corresponding
confined reciprocity. The resulting geometry-dependent Seebeck and Peltier
responses therefore obey the Kelvin relation.

The comparison between linear and parabolic dispersions reveals that the size dependence of these thermoelectric coefficients can arise from fundamentally different microscopic mechanisms. For a linear band, the energy-independent carrier velocity leaves the thermopower unchanged in the resistive limit of the Callaway model. When momentum-conserving and resistive relaxation times differ, the common geometric factorization of the electrical and thermoelectric responses is broken by the collective momentum response, producing a finite-width correction to the thermopower. For a parabolic band, the energy dependence of boundary propagation modifies the thermopower already in the resistive limit, while its additional dependence on the balance between momentum-conserving and resistive collisions is comparatively weak. In both cases, the finite-size correction reduces the Seebeck coefficient relative to its bulk value. The finite-size thermoelectric response therefore retains kinetic information beyond that captured by the hydrodynamic length scale alone.

The constitutive closure is supported by two complementary benchmarks.
For electronic heat transport, the resulting Guyer--Krumhansl-type equation
quantitatively accounts for size-dependent thermal-transport measurements
in Cu and Ni in both planar films and three-dimensional porous geometries.
For the coupled thermoelectric problem, direct comparison with the
spatially resolved Boltzmann equation provides a microscopic benchmark
of the finite-width response.

The Cu and Ni measurements were described using an effectively decoupled
thermal FDF, whereas the feedback of the local electrical counterflows on the averaged
heat current is second order in the thermoelectric coupling. For the
degenerate electron systems considered here, the relative scale of this feedback is
$ZT_e=S^2\sigma T/\kappa_e=O[(T/T_F)^2]$, so it lies beyond the
Sommerfeld order retained in the thermal benchmark. These measurements
therefore provide an experimental benchmark of the thermal sector of the
coupled FDF.

The Boltzmann benchmark establishes the quantitative range of the
constitutive closure. The FDF captures the finite-width response already
for characteristic dimensions comparable to $v_F\tau_C$, with an accuracy
that improves systematically as the width increases. The remaining
deviations at the smallest widths reflect higher kinetic moments generated
by streaming beyond the retained flux space, identifying the fast kinetic
propagation length, rather than the long hydrodynamic scale alone, as the
relevant control scale for the closure. This leaves an intermediate-width
regime in which finite-size thermoelectric effects remain substantial
while the FDF already provides a quantitatively useful description.

The geometry dependence of the Seebeck and Peltier coefficients 
suggests experimental configurations to further validate and explore the implications of the nonlocal thermoelectric phenomena predicted by the FDF. In particular, we presented a differential strategy in which two branches of the same material share
the same bulk thermoelectric response but have different transverse
dimensions. Making one branch macroscopic and the other comparable to a
kinetic length cancels the common bulk contribution and isolates the
geometry-dependent Seebeck signal; the reciprocal electrical configuration
probes the corresponding Peltier response.

These results also place the present constitutive equations in the context
of other kinetic descriptions of nonlocal thermoelectric transport
\cite{Battiato2017,Coulter2025}. A distinctive feature of the present
construction is that charge and heat currents are retained as independent
constitutive variables. The relation to descriptions based on a single
momentum mode is particularly transparent for a parabolic dispersion:
because the charge current is directly proportional to momentum density, retaining
the momentum mode amounts to retaining the charge current $\mathbf j$
as the single constitutive flux. The two-current closure instead retains
both $\mathbf j$ and the heat flux $\mathbf q$, preserving the spatial dynamics of the
heat current and the coupled thermoelectric response within the same
kinetic description. This distinction also extends to finite geometries:
retaining both physical currents preserves their boundary response
explicitly and allows the kinetic boundary conditions and bulk
constitutive equations to be formulated in terms of the same fields.

The explicit evaluation presented here is carried out in the degenerate
regime. Electron--phonon scattering enters through resistive relaxation
rather than through an independent nonequilibrium phonon distribution.
Extending the explicit evaluation beyond the degenerate regime and
developing a coupled electron--phonon kinetic description within the FDF
are natural directions for future work. The latter extension would also
provide a setting for generalizing the flux-based construction to
additional interacting transport channels.

The broader implication is that kinetic nonlocality need not be treated
only at the level of the full distribution function: it can be encoded
in closed constitutive equations for experimentally accessible fluxes,
with boundary conditions derived from the underlying kinetics and
expressed in terms of the same physical variables. Solving the resulting
boundary-value problem determines the complete geometry-dependent
transport matrix, allowing substantial finite-size modifications of the
electrical, thermal, and thermoelectric response to be described within
a single framework. In this sense, the FDF provides a direct bridge
between the microscopic Boltzmann equation and measurable transport
in finite geometries.

\section*{Acknowledgments}
This work is supported by the Spanish Ministerio de
Ciencia, Innovacion y Universidades under Grants No.
PID2021-122322NB-I00 and No. TED2021-129612B-C22
(MCIU/AEI/10.13039/501100011033/FEDER UE) and the
AGAUR-Generalitat de Catalunya under Grant No. 2021-
SGR-00644.

\appendix

\section{Detailed FDF derivation of the nonlocal constitutive equation for isothermal charge transport}
\label{app:isothermal_derivation}

In this appendix, we use the Flux Derivative Formalism to derive the 
nonlocal equation for the current density, Eq. \eqref{GKE} in the main text. 
The derivation follows systematic steps, outlined below for isotropic 
systems under uniform temperature.
For completeness, we reproduce the relevant equations from the main text so that the appendix is largely self-contained.

\subsection*{Step (i): Constitutive expansion of the distribution function}

The nonequilibrium distribution function is written as a constitutive expansion 
in terms of the retained flux (charge current $\mathbf j$) and its spatial and 
temporal derivatives:
\begin{equation}
	\label{ansatz_ap}
   f=f^{0}+\vec{\beta}_j\cdot \mathbf j+\vec{\gamma}_j \cdot \frac{\partial \mathbf j}{\partial t}+\overset{=}{G}_{j} : {\nabla}\mathbf j,
\end{equation}
where $\vec{\beta}_j$, $\vec{\gamma}_j$ and
$\overset{=}{G}_{j}$ are the microscopic weight functions associated with the retained variables. From now on in this section, we drop the subindex $j$ in the weight functions.
We restrict the discussion to isotropic materials. In this case one has
$\overset{=}{G}\equiv \overset{=}{G}_{1}+\overset{=}{G}_{2}$, with
$\overset{=}{G}_{1}\equiv g_{1}\hat{k} \hat{k}$ and
$\overset{=}{G}_{2}\equiv g_{2}\mathbb{I}$, where $\hat{k}$ denotes the unit
vector along $\vec{k}$, $\mathbb{I}$ is the identity tensor, and $g_{1}$ and
$g_{2}$ are scalar functions of $\vec{k}$; the symbol ``:'' denotes double
tensor contraction. Hereafter in this section, we omit the subscript $j$ identifying the current density in $\vec{\beta}_j$, $\vec{\gamma}_j$ and
$\overset{=}{G}_{jj}$; all remaining subscripts denote spatial coordinates.

From the definition of 
charge current and the constitutive expansion:
\begin{equation}
	\label{j_def}
	\mathbf j=\int e \vec{v} f d\lambda=
    \int e \vec{v} \vec{\beta}  d\lambda \cdot \mathbf j+\int e \vec{v} \vec{\gamma}  d\lambda \cdot \frac{ \partial \mathbf j}{\partial t},
\end{equation}
we obtain consistency conditions on the weight functions:
\begin{equation}
\label{consistency_beta_app}
\int e \vec{v} \vec{\beta} d\lambda = \mathbb{I},
\end{equation}
\begin{equation}
\label{consistency_gamma_app}
\int  \vec{v} \vec{\gamma} d\lambda = 0.
\end{equation}
The integral of $\overset{=}{G}_{j}$ identically vanishes by parity in the isotropic systems under study.

\subsection*{Step (ii): Transport equations by projection of the Boltzmann equation and closure conditions}

Projecting the Boltzmann equation to charge and current density gives the 
transport equations:
\begin{equation}
    		e\frac{\partial n}{\partial t}+{\nabla}\cdot \mathbf j=0,
            \label{charge_conservation_app}
\end{equation}
\begin{equation}
\label{j_flux}
	\frac{\partial \mathbf j}{ \partial t}+{\nabla}\cdot \overset{=}{\Pi} + \int e \vec{v} \, \vec{F} \cdot  {\nabla}_{\vec{p}} f=\int e  \vec{v} \, C(f)d\lambda,
\end{equation}
with $\overset{=}{\Pi}=\int e \vec{v}\vec{v} f$ the flux of current density. By keeping just linear terms, the prefactor of the the force field $\vec{F}$ yields $\int e \vec{v}\vec{v} \,\partial_{\epsilon} f^{0}$. 

Substituting Eq. \eqref{ansatz_ap} into Eqs. \eqref{j_flux} and 
\eqref{charge_conservation_app}, we close the system. The current flux becomes:
\begin{equation}
	\overset{=}{\Pi}= \int e \vec{v} \vec{v} f^{0} d\lambda + \int e \vec{v} \vec{v} \overset{=}{G} d\lambda : {\nabla}\mathbf j,
\end{equation}
and the collision term yields:
\begin{equation}
\int e \vec{v} C(f) d\lambda=
\int e \vec{v} C(\vec{\beta}) d\lambda \cdot \mathbf j+ \int e \vec{v}_\lambda C(\vec{\gamma}) d\lambda \cdot \frac{\partial \mathbf j}{\partial t}.
\end{equation}

Substituting these into the projected equation gives:
\[
 \frac{\partial \mathbf j}{\partial t} + \int e v_x v_x \,\partial_{\epsilon} f^{0} d\lambda \, \left(\vec{F} - \nabla \mu\right) + \int e v_x v_y G_{xy} \nabla^2 \mathbf j+   
\]
\[
+ \left(\int e v_x^2 G_{xx} d\lambda -\int e v_x v_y G_{xy} d\lambda \right) \nabla(\nabla   \cdot \mathbf j)= 
\]
\[
=\int e v_x C(\beta_x) d\lambda \cdot \mathbf j+ \int e v_x C(\gamma_x) d\lambda \cdot \frac{\partial \mathbf j}{\partial t}.
\]

The above equation can be rewritten in closed form as:
\begin{equation}
\label{GKE_app}
\tau\frac{\partial \mathbf j}{\partial t}+\mathbf j=\sigma \left(\vec{E} - \frac{\nabla \mu}{e} \right)+\ell^2\left(\nabla^2\mathbf j+\alpha \nabla(\nabla\cdot \mathbf j)\right), 
\end{equation}
by identifying the transport parameters $\sigma, \tau, \ell$ and $\alpha$ as defined 
in Eqs. \eqref{sigma}-\eqref{alpha}. This is the generalized nonlocal equation 
for charge transport (Eq. \eqref{GKE} in the main text).

\subsection*{Step (iii): Determination of the weight functions equations}

To obtain the equations for the microscopic coefficents $\vec{\beta}$, $\vec{\gamma}$, 
$\overset{=}{G}_{1}$, and $g_{2}$, we substitute expansion  \eqref{ansatz_ap} into the 
Boltzmann equation itself. The goal now is to express this equation in terms of the retained macroscopic variables $\mathbf j$ and its first derivatives. We use the transport equations \eqref{charge_conservation_app} and 
\eqref{GKE_app} to eliminate $\frac{\partial n}{\partial t}$ and $\vec{E}$, and neglect second-order derivatives to get:
\begin{align}
\begin{split}
\label{BTE_GK}
	\frac{1}{e} \left(\frac{\partial \mu}{\partial n}\right)_{T} \! \partial_{\epsilon} f^{0} \,
\nabla\cdot \mathbf j +\vec{v} \vec{\beta} : \nabla\mathbf j\\
    +\frac{e}{\sigma} \vec{v} \, \partial_{\epsilon} f^{0} \! \cdot \mathbf j+\frac{e\tau}{\sigma} \vec{v} \,\partial_{\epsilon} f^{0} \cdot \frac{\partial\mathbf j}{\partial t} + \vec{\beta} \cdot \frac{\partial\mathbf j}{\partial t}\\
    = C(\vec{\beta})\cdot \mathbf j+ C(\vec{\gamma})\cdot\frac{\partial \mathbf j}{\partial t} +C(\overset{=}{G}_{1}):\nabla\mathbf j  + C(g_{2}) \nabla\cdot \mathbf j  
\end{split}
\end{align}

Finally, identifying coefficients of the independent macroscopic variables ($\mathbf j$, 
$\frac{\partial \mathbf j}{\partial t}$, $\nabla\mathbf j$, $\nabla\cdot\mathbf j$) 
yields the equations for the weights $\vec{\beta}$, $\vec{\gamma}$, 
$\overset{=}{G}_{1}$, and $g_{2}$, as given in Eqs. \eqref{beta}-\eqref{g2}. The 
independence of these variables is guaranteed by the form of Eq. \eqref{GKE_app} 
for the current density.

\section{Callaway collision model}
\label{app:callaway_singleflux}
We apply the FDF to the Callaway model for isotropic systems and obtain the expressions grasped in the main text. In this model, the collision operator is
\begin{equation}
    C(f)=- \frac{f-f^0}{\tau_R}-  \frac{f-f^{disp}}{\tau_{ee}}
    \label{Callaway_A}
\end{equation}
where $\tau_{R}$ is the average collision time for resistive collisions (those not conserving crystal momentum), and $\tau_{ee}$ the one corresponding to electron-electron scattering (which do conserve momentum); both are mode-independent. The displaced distribution close to equilibrium writes 
\begin{equation}
f^{disp}=f^0 - \vec{p} \!\cdot \! \vec{u}\, \partial_{\epsilon} f^{0}
\label{disp_A}
\end{equation}
with $\vec{u}$ a velocity parameter related to the momentum density. Its value is determined by imposing that momentum is conserved in electron-electron scattering
\begin{equation}
\label{pcons}
\int \vec{p} \, \frac{f-f^{disp}}{\tau_{ee}}=0.    
\end{equation}

This leads to $\vec{u}=\frac{\int \vec{p} f d\lambda}{\int p_x^2 (-\partial_{\epsilon} f^{0} d\lambda)}$, which depends on the distribution function $f$. If one uses ansatz \eqref{ansatz_f}, one obtains
\begin{equation}
	\vec{u}= A \mathbf j+ B \frac{\partial \mathbf j}{\partial t},
 \label{u}
\end{equation}
with 
\begin{equation}
	A =\frac{\int p_x \beta_x d\lambda}{\int p_x^2 (-\partial_{\epsilon} f^{0} d\lambda)}, \hspace{5mm} B =\frac{\int p_x \gamma_x d\lambda}{\int p_x^2 (-\partial_{\epsilon} f^{0} d\lambda)}. 
 \label{AB}
\end{equation}

Introducing Eqs. \eqref{disp_A}, \eqref{u} and ansatz \eqref{ansatz_ap} into the collision operator \eqref{Callaway_A} yields the specific expressions for the Callaway model 
\begin{subequations}
\begin{align}
	C(\vec{\beta}) &= -\frac{\vec{\beta}}{\tau_C} +A  \frac{\vec{p}}{\tau_{ee}} \partial_{\epsilon} f^{0} \label{Callaway_beta} \\
 C(\vec{\gamma}) &=- \frac{\vec{\gamma}}{\tau_C} +B  \frac{\vec{p}}{\tau_{ee}} \partial_{\epsilon} f^{0}  \label{Callaway_gamma}\\ C(\overset{=}{G})&=- \frac{\overset{=}{G}}{\tau_C},
 \label{Callaway_G}
\end{align}
\end{subequations}
with $\tau_C^{-1} \equiv \tau_R^{-1}+\tau_{ee}^{-1}$
the average collision time. By introducing this expressions in the general equations for the weights, Eqs. \eqref{beta}-\eqref{g2}, we find for Callaway model
\begin{subequations}
\begin{align}
	\vec{\beta} &= - \tau_C \partial_{\epsilon} f^{0} \left( \frac{e}{\sigma} \vec{v} +  \frac{A}{\tau_{ee}} \vec{p} \right) \label{beta_cal}\\
 \vec{\gamma} &= 0 \\
 \overset{=}{G}_1 &= - \tau_C \, \vec{v}\vec{\beta}  \label{g1_cal} \\
 g_2 &= - \frac{\tau_C}{e} \left(\frac{\partial \mu}{\partial n}\right)_{T} \partial_{\epsilon} f^{0}
\label{g2_cal}
\end{align}
\end{subequations}
 One can introduce expression \eqref{beta_cal} into \eqref{AB} to obtain parameter $A=\frac{e\tau_R}{\sigma}\frac{v_F}{p_F}$, so that $\vec{\beta}$ simplifies to
\begin{equation}
  \vec{\beta} =  - \frac{e\tau_R}{\sigma} \vec{v} \partial_{\epsilon} f^{0}.
  \label{beta_cal2}  
\end{equation}
This implies that $C(\vec{\beta})=-\vec{\beta}/\tau_R$, meaning that in the distribution function the contribution related to the current density relaxes just through resistive collisions, as expected since electron-electron scattering does not relax momentum. In contrast, Eqs. \eqref{g1_cal}-\eqref{g2_cal} show that the part of the distribution related to the flux of current ($\overset{=}{G}$) relaxes through both scattering types. They rewrite as 
\begin{equation}
     \overset{=}{G} = \frac{e \tau_R \tau_C}{\sigma} \partial_{\epsilon} f^{0} (\vec{v} \vec{v} - \frac{v^2}{d} \mathbb{I}),
\end{equation}
which is a traceless tensor, indicating that there is no bulk electronic viscosity in the Callaway model.

The result $\vec{\gamma} = 0$ can be obtained from constrain \eqref{consistency_gamma_app}. 
For a sharp Fermi surface one has $\vec{\gamma}\simeq \vec{\gamma}(p_F) \partial_{\epsilon} f^0$, so that
\[
\int v_x \gamma_x \simeq \int v_x \gamma_x(p_F) \partial_{\epsilon} f^0 \simeq \frac{v_F \gamma(p_F)}{d} \int \partial_{\epsilon} f^0=0, 
\]
so that $\gamma(p_F)=0$ and $\vec{\gamma}=0$.
%One has $\vec{\gamma} = \vec{\gamma'} \partial_{\epsilon} f^0$, so that for a sharp Fermi surface 
%\[ \int v_x \gamma_x = \int v_x \gamma'_x \partial_{\epsilon} f^0 \simeq \frac{v_F \gamma'_F}{d} \int \partial_{\epsilon} f^0=0, \]

\vspace{1cm}
Transport coefficients are obtained from the general expressions \eqref{sigma}-\eqref{alpha} and the consistency equations. For isotropic dispersion relations in $d$-dimensional system they yield 
\begin{subequations}
\begin{align}
	\sigma &=e^2 \tau_R \int v_x^2 (-\partial_\epsilon f^0) \, d\lambda \\
  \ell^2 &= \frac{1}{d+2}  \tau_C \tau_R v_F^2 \\
 \alpha &= \frac{d-2}{d}. 
 \label{coefs_Callaway}
\end{align}
\end{subequations}

\subsection{Thermal single-flux resistive limit}
\label{app:thermal_rta_weights}

For the thermal benchmark of Sec.~\ref{sec:thermal_validation}, the retained
flux is the electronic heat current
\begin{equation}
 \mathbf q=\int (\epsilon-\mu)\vec v\,f\,d\lambda .
\end{equation}
In the relaxation-time approximation, $C(f)=-(f-f^0)/\tau_R$, the same
single-flux derivation used above gives a local weight proportional to
$\tau_R(\epsilon-\mu)\vec v$ and a first-gradient weight obtained by one
additional action of the streaming operator.  For a transverse variation,
the coefficient of $\nabla^2\mathbf q$ can therefore be written as the ratio
\begin{equation}
 \ell_q^2=\tau_R^2
 \frac{\displaystyle\int (\epsilon-\mu)^2 v_x^2v_y^2
 (-\partial_\epsilon f^0)\,d\lambda}
 {\displaystyle\int (\epsilon-\mu)^2 v_x^2
 (-\partial_\epsilon f^0)\,d\lambda} .
 \label{eq:ellq_ratio_app}
\end{equation}
In the degenerate limit the velocity can be evaluated at the Fermi surface.
For an isotropic $d$-dimensional system,
$\langle v_x^2\rangle=v_F^2/d$ and
$\langle v_x^2v_y^2\rangle=v_F^4/[d(d+2)]$.
Equation~\eqref{eq:ellq_ratio_app} then gives
\begin{equation}
 \ell_q^2=\frac{v_F^2\tau_R^2}{d+2},
 \label{eq:ellq_RTA_app}
\end{equation}
which is Eq.~\eqref{ellq_RTA} of the main text.  This is also the kinetic
limit of the Callaway result, $\tau_C\to\tau_R$.

\section{Degenerate isotropic evaluation of the nonlocal matrix}
\label{app:M_isotropic}

We now evaluate Eq.~\eqref{eq:M_general} for the isotropic degenerate bands considered in Sec.~\ref{sec:degenerate_isotropic}. The matrix $M$ multiplying the Laplacian term retained in the main text is obtained from a transverse configuration, with the current along $x$ and the spatial variation along $y$. Isotropy makes the resulting coefficient independent of the particular choice of transverse directions.

For the Callaway operator, Eq.~\eqref{eq:callaway_R} may be written as
\begin{equation}
R=\tau_C I+(\tau_R-\tau_C)\mathcal P .
\label{eq:R_Callaway_app}
\end{equation}
Here $\mathcal P$ is the projector onto the momentum subspace.
For either charge or heat current, the vectors in $A_x$ are proportional to $v_x$ times a function of energy. The projection $\mathcal P V_yRA_x$ therefore contains angular averages of the form $\langle p_i v_yv_x F(\epsilon)\rangle$. These averages vanish by inversion symmetry, and hence
\begin{equation}
\mathcal P V_yRA_x=0.
\label{eq:PVR_zero_app}
\end{equation}
It follows that
\begin{equation}
RV_yRA_x=\tau_CV_yRA_x .
\label{eq:RVRA_app}
\end{equation}
Using Eq.~\eqref{eq:M_general}, the transverse nonlocal matrix is therefore
\begin{equation}
M=\tau_C LB^{-1}DL^{-1},
\qquad
D\equiv A_x^TV_y^2RA_x .
\label{eq:M_Callaway_app}
\end{equation}

To evaluate the remaining matrices, write $\xi=\epsilon-\mu$ and introduce
\begin{equation}
t_2=\frac{\pi^2}{3}(k_BT)^2 .
\label{eq:t2_app}
\end{equation}
The charge- and heat-current weights are proportional to $e v_x$ and $\xi v_x$, respectively, while the momentum weight can be expressed as
\begin{equation}
p_x=b(\epsilon)v_x .
\label{eq:px_bvx_app}
\end{equation}
For convenience, we recall the spectral functions introduced in Sec.~\ref{sec:degenerate_isotropic},
\[
a(\epsilon)=\frac{D(\epsilon)v^2(\epsilon)}{d},\qquad
b(\epsilon)=\frac{p(\epsilon)}{v(\epsilon)},\qquad
c(\epsilon)=\frac{v^2(\epsilon)}{d+2}.
\]
With these definitions, isotropy gives
\begin{subequations}\label{eq:isotropic_averages_app}
\begin{align}
\langle v_x^2 F(\epsilon)\rangle
&=\int d\epsilon\,[-f_0'(\epsilon)]\,a(\epsilon)F(\epsilon),\\
\langle v_x^2v_y^2 F(\epsilon)\rangle
&=\int d\epsilon\,[-f_0'(\epsilon)]\,a(\epsilon)c(\epsilon)F(\epsilon).
\end{align}
\end{subequations}
{\color{black}
Because the heat-sector entries start at $O(t_2)$, their inverses contain
$1/t_2$.  Therefore a relative $O[(k_BT/\mu)^2]$ evaluation of the
nonlocal matrix requires the thermal moments one Sommerfeld order beyond
the leading term.  The required moments are
\begin{subequations}\label{eq:Sommerfeld_app}
\begin{align}
\int d\epsilon\,[-f_0']F
&=F_F+\frac{t_2}{2}F_F''+\frac{7t_2^2}{40}F_F''''+O(T^6),\\
\int d\epsilon\,[-f_0']\,\xi F
&=t_2F_F'+\frac{7t_2^2}{10}F_F'''+O(T^6),\\
\int d\epsilon\,[-f_0']\,\xi^2F
&=t_2F_F+\frac{21t_2^2}{10}F_F''+O(T^6).
\end{align}
\end{subequations}
The $t_2^2$ terms do not contribute to the leading matrix displayed in the
main text, but they must be retained whenever the relative $O(T^2)$
corrections to its diagonal entries, or combinations containing them, are
formed.}

It is useful to display the matrix algebra explicitly.  With
\(
A_x=(e v_x,\,\xi v_x)
\), the current-overlap matrix is, to the order needed below,
\begin{equation}
B=
\begin{pmatrix}
e^2a & e t_2 a'\\
e t_2 a' & t_2 a
\end{pmatrix}_{\!F}
+\Delta B,
\label{eq:B_Sommerfeld_app}
\end{equation}
{\color{black}where $\Delta B$ contains the $O(T^2)$ Sommerfeld correction to the
upper-left element and the higher-order thermal moments specified in
Eq.~\eqref{eq:Sommerfeld_app}.  They are not needed to obtain the leading
matrix below, but they do not in general cancel from the relative
$O(T^2)$ corrections to the diagonal elements.}

The momentum projection entering $R$ is determined by
\begin{equation}
\begin{aligned}
\langle P_x|A_x\rangle
&=
\begin{pmatrix}
eab & t_2(ab)'
\end{pmatrix}_{\!F}
+O(T^4),\\
\langle P_x|P_x\rangle
&=(ab^2)_F+O(T^2).
\end{aligned}
\label{eq:momentum_overlaps_app}
\end{equation}
Consequently $L=A_x^TRA_x$ has the off-diagonal element
\begin{equation}
L_{12}=L_{21}
=e t_2\left[
\tau_R a'
+(\tau_R-\tau_C)a\frac{b'}{b}
\right]_{\!F}+O(T^4),
\label{eq:L12_app}
\end{equation}
while its leading diagonal terms are
\begin{equation}
L_{11}=e^2\tau_R a_F+O(T^2),
\qquad
L_{22}=t_2\tau_C a_F+O(T^4).
\label{eq:Ldiag_app}
\end{equation}

The remaining matrix in Eq.~\eqref{eq:M_Callaway_app} is
$D=A_x^TV_y^2RA_x$.  Using
$\langle A_x|V_y^2|P_x\rangle$ together with
Eq.~\eqref{eq:isotropic_averages_app} gives
\begin{subequations}\label{eq:D_elements_app}
\begin{align}
D_{11}&=e^2\tau_R(ac)_F+O(T^2),\\
D_{22}&=t_2\tau_C(ac)_F+O(T^4),
\end{align}
\begin{align}
D_{12}={}&e t_2\Bigg[
\tau_C(ac)' \\[-1mm]
&+(\tau_R-\tau_C)ac
\left(\frac{a'}a+\frac{b'}b\right)
\Bigg]_{\!F}+O(T^4),
\end{align}
\begin{align}
D_{21}={}&e t_2\Bigg[
\tau_C(ac)' \\[-1mm]
&+(\tau_R-\tau_C)ac
\left(\frac{a'}a+\frac{b'}b+\frac{c'}c\right)
\Bigg]_{\!F}+O(T^4).
\end{align}
\end{subequations}
Here $D$ need not be symmetric because $V_y^2$ and $R$ do not in
general commute.

Substituting Eqs.~\eqref{eq:B_Sommerfeld_app}--\eqref{eq:D_elements_app}
into
\(
M=\tau_C LB^{-1}DL^{-1}
\), expanding consistently through the retained Sommerfeld order, and
only then simplifying, gives
\begin{equation}
M=
\begin{pmatrix}
\tau_R\tau_C c & M_{12}\\[1mm]
M_{21} & \tau_C^2c
\end{pmatrix}_{\!F},
\label{eq:M_general_tau_app}
\end{equation}
with
\begin{subequations}\label{eq:M12M21_tau_app}
\begin{align}
M_{12}&=e\tau_C\left[\tau_Rc'+(\tau_R-\tau_C)c\frac{b'}{b}\right]_{\!F},\\
M_{21}&=\frac{t_2\tau_C}{e}\left[\tau_Cc'+(\tau_R-\tau_C)c\left(\frac{a'}{a}+\frac{b'}{b}\right)\right]_{\!F}.
\end{align}
\end{subequations}
The derivatives of $a$ that enter $B$, $L$, and $D$ cancel from
$M_{12}$, whereas $M_{21}$ retains the combination
$a'/a+b'/b$.  {\color{black}At leading degenerate order} the diagonal entries reduce to
$M_{11}=\tau_R\tau_Cc_F$ and
$M_{22}=\tau_C^2c_F$. {\color{black}Their relative $O[(k_BT/\mu)^2]$
corrections are generally nonzero.  They are not part of the leading
matrix used in Sec.~\ref{sec:degenerate_isotropic}; when the open-circuit
thermal equation is reduced consistently, they must be combined with the
thermoelectric terms before the Sommerfeld truncation is made, as shown in
Appendix~\ref{app:thermal_benchmark}.} Finally, using
$r=\tau_C/\tau_R$ and $\ell_C^2=\tau_C^2c_F$ yields
Eqs.~\eqref{eq:M_general_isotropic} and \eqref{eq:mij_general} of the
main text.

For a power-law dispersion,
$\epsilon=A p^z$, one just substitutes
\begin{equation}
\frac{a'}{a}=\frac{d+z-2}{z\epsilon},\qquad
\frac{b'}{b}=\frac{2-z}{z\epsilon},\qquad
\frac{c'}{c}=\frac{2(z-1)}{z\epsilon}.
\label{eq:abc_power_law}
\end{equation}

\section{Onsager reciprocity}
\label{app:bulk_reciprocity}

In the absence of a magnetic field, Onsager reciprocity for a spatially
nonlocal constitutive relation requires
\begin{equation}
K(\mathbf k)=K^T(-\mathbf k),
\label{eq:Onsager_general_k}
\end{equation}
where $\mathbf k$ is the wave vector conjugate to the spatial coordinate.
For the isotropic systems considered here, the response is even under
$\mathbf k\rightarrow-\mathbf k$, so that reciprocity reduces to
$K(\mathbf k)=K^T(\mathbf k)$.

We now verify that the reduced two-flux constitutive law preserves this
property to the Sommerfeld order retained in the main text. In real
space, the stationary FDF equations can be written as
\begin{equation}
\mathbf J=L\mathbf X+M\nabla^2\mathbf J .
\label{eq:FDF_real_onsager}
\end{equation}
Upon Fourier transformation,
$\nabla\rightarrow i\mathbf k$ and $\nabla^2\rightarrow-k^2$, with
$k=|\mathbf k|$, this becomes
\begin{equation}
\mathbf J(\mathbf k)=K(k)\mathbf X(\mathbf k),
\qquad
K(k)=(I+k^2M)^{-1}L .
\label{eq:K_fourier_onsager}
\end{equation}
Since the local kinetic matrix
\begin{equation}
L=A^T R A
\end{equation}
is symmetric, the only nontrivial condition concerns the nonlocal matrix
$M$. From Eq.~\eqref{eq:K_fourier_onsager},
\begin{equation}
K^{-1}(k)=L^{-1}+k^2L^{-1}M.
\label{eq:Kinv_onsager}
\end{equation}
Thus $K(k)$ is reciprocal if
\begin{equation}
H\equiv L^{-1}M=H^T,
\label{eq:H_symmetric}
\end{equation}
or, equivalently, if $ML$ is symmetric.

For the transverse geometry used above, Eq.~\eqref{eq:M_Callaway_app} gives
\begin{equation}
ML=\tau_C L B^{-1} A^T V_y^2 R A.
\label{eq:ML_start}
\end{equation}
The action of $R$ on the two current current subspace can always be decomposed as
\begin{equation}
RA=AS+\chi,
\qquad A^T\chi=0,
\label{eq:RA_decomp}
\end{equation}
where $AS$ is the projection of $RA$ onto the two-dimensional current subspace and $\chi$ is the orthogonal remainder.  Multiplying Eq.~\eqref{eq:RA_decomp} by $A^T$ gives
\begin{equation}
L=A^TRA=BS,
\qquad S=B^{-1}L.
\label{eq:S_def}
\end{equation}
Because both $B$ and $L$ are symmetric, $LB^{-1}=S^T$.  Equation~\eqref{eq:ML_start} may therefore be written as
\begin{equation}
ML=\tau_C S^T A^T V_y^2(AS+\chi).
\label{eq:ML_decomp}
\end{equation}
Defining the symmetric matrix
\begin{equation}
D_0\equiv A^T V_y^2 A=D_0^T,
\end{equation}
we obtain
\begin{equation}
ML=\tau_C S^T D_0 S
   +\tau_C S^T A^T V_y^2\chi.
\label{eq:ML_two_terms}
\end{equation}
For a general degenerate isotropic band, the second term in
Eq.~\eqref{eq:ML_two_terms} is $O(T^4)$.  This follows directly from the
Sommerfeld expansion.  The two rows of $A^T$ have energy dependences $1$ and $\xi$, with
$\xi=\varepsilon-\mu$.  Thus the condition $A^T\chi=0$ states that the
constant and linear in $\xi$ projections of $\chi$ vanish.

After the angular average, the additional factor $V_y^2$ entering
$A^T V_y^2\chi$ is
\begin{equation}
 c(\varepsilon)=\frac{v^2(\varepsilon)}{d+2}
 =c_F+c_F'\xi+\frac{1}{2}c_F''\xi^2+\cdots .
\end{equation}
For the first row of $A^T V_y^2\chi$, the terms proportional to $c_F$
and $c_F'\xi$ are precisely the constant and linear projections already
removed by $A^T\chi=0$.  For the second row, the $c_F$ term is the
linear projection and vanishes for the same reason.  The remaining
terms contain at least two additional powers of $\xi$.  Since the
orthogonal remainder $\chi$ itself starts at quadratic order in the
energy expansion, their first nonvanishing Sommerfeld contribution is
of fourth order.  Hence
\begin{equation}
A^T V_y^2\chi=O(T^4).
\label{eq:chi_order}
\end{equation}
Consequently,
\begin{equation}
ML=\tau_C S^T D_0 S+O(T^4)
\label{eq:ML_symmetric}
\end{equation}
is symmetric through the retained $O(T^2)$ order.  Equations~\eqref{eq:H_symmetric} and \eqref{eq:ML_symmetric} then imply
\begin{equation}
K^T(k)=K(k)+O(T^4).
\end{equation}
This establishes bulk Onsager reciprocity for an arbitrary isotropic degenerate band to the order used in the FDF coefficients.

For the power-law dispersions $z=1$ and $z=2$ considered explicitly in the paper, $RA$ closes on the current subspace, so that $\chi=0$.  In these two cases Eq.~\eqref{eq:ML_symmetric} contains no remainder and the bulk reciprocity relation is exact within the energy-independent Callaway model.  The corresponding statement for the confined problem requires, in addition, compatibility of the kinetic boundary condition with the same metric $H$; this is shown in Appendix~\ref{app:boundary_conditions}.

\section{Boundary conditions for coupled charge and heat transport}
\label{app:boundary_conditions}

\subsection{Kinetic boundary condition}

We derive here the kinetic boundary condition for the coupled FDF at a diffusely scattering surface. At each point of the boundary we choose the local normal coordinate $n$ to point from the material towards the wall, so that $v_n>0$ for incident states and $v_n<0$ for outgoing states, and denote integration over the outgoing half space by $[\,\cdot\,]_-$.

The FDF distribution at the wall is
\begin{equation}
\phi_w=RA L^{-1}\mathbf J_w-RV_nRA L^{-1}\partial_n\mathbf J_w,
\label{eq:phi_wall_app}
\end{equation}
where $V_n$ denotes multiplication by the normal velocity component. For a fully diffuse wall, carriers leaving the wall are re-emitted in local equilibrium, so that their nonequilibrium correction vanishes, $\phi_w^-=0$. Since the FDF retains only the charge and heat currents as nonequilibrium variables, we impose this condition on the corresponding half-range flux moments, $[A^T |V_n|\phi_w]_- =0$. Substituting Eq.~\eqref{eq:phi_wall_app} gives
\begin{equation}
\begin{split}
[A^T |V_n|RA]_-L^{-1}\mathbf J_w
&-
[A^T |V_n|RV_nRA]_-L^{-1}\\
&\hspace{18mm}\times\partial_n\mathbf J_w=0.
\end{split}
\label{eq:wall_half_range}
\end{equation}
Defining
\begin{equation}
{\cal A}_w=[A^T |V_n|RA]_-L^{-1},\qquad
{\cal B}_w=-[A^T |V_n|RV_nRA]_-L^{-1},
\label{eq:Aw_Bw_app}
\end{equation}
the boundary condition becomes
\begin{equation}
{\cal A}_w\mathbf J_w=-{\cal B}_w\partial_n\mathbf J_w.
\label{eq:wall_general_matrix}
\end{equation}
If ${\cal A}_w$ is invertible, this becomes
\begin{equation}
\mathbf J_w=-\Lambda\,\partial_n\mathbf J_w,
\qquad
\Lambda={\cal A}_w^{-1}{\cal B}_w.
\label{eq:Lambda_general_app}
\end{equation}
Thus, in the coupled problem the slip length is in general a $2\times2$ matrix in charge--heat space.

For the Callaway operator,
\begin{equation}
R=\tau_C I+(\tau_R-\tau_C)\mathcal P.
\end{equation}
For an isotropic system, $RA$ is even under reversal of the normal momentum whereas $V_nRA$ is odd. Consequently
\begin{equation}
\mathcal P V_nRA=0,
\end{equation}
and hence
\begin{equation}
RV_nRA=\tau_CV_nRA.
\label{eq:RVnRA_wall}
\end{equation}
Defining
\begin{equation}
W_0=[A^T |V_n|RA]_-,
\qquad
W_1=[A^T |V_n|^2RA]_-,
\label{eq:W0W1_wall}
\end{equation}
Eq.~\eqref{eq:wall_general_matrix} becomes
\begin{equation}
W_0L^{-1}\mathbf J_w
=-\tau_CW_1L^{-1}\partial_n\mathbf J_w,
\end{equation}
and the Callaway boundary condition is therefore
\begin{equation}
\mathbf J_w=-\Lambda_C\partial_n\mathbf J_w,\qquad
\Lambda_C=\tau_C L W_0^{-1}W_1L^{-1}.
\label{eq:Lambda_Callaway_app}
\end{equation}
The matrix ordering in Eq.~\eqref{eq:Lambda_Callaway_app} is essential in the coupled problem.

For a linear band, $z=1$, this result is in fact exact within the Callaway model with energy-independent relaxation times. Since the magnitude of the velocity is independent of energy, the energy dependence of the matrices entering Eq.~\eqref{eq:Lambda_Callaway_app} cancels, giving for arbitrary $r=\tau_C/\tau_R$
\begin{equation}
\begin{aligned}
\Lambda_C^{2D}(z=1,r)&=\frac{3\pi}{16}v_F\tau_C I,\\
\Lambda_C^{3D}(z=1,r)&=\frac{8}{15}v_F\tau_C I.
\end{aligned}
\label{eq:Lambda_z1_app}
\end{equation}
There are therefore no off-diagonal boundary couplings or Sommerfeld corrections in the linear-band case.

For a parabolic band, $z=2$, the first Sommerfeld corrections generate off-diagonal boundary couplings. To the order required for the thermoelectric strip problem,
\begin{equation}
\Lambda_C^{2D}=\frac{3\pi}{16}v_F\tau_C
\begin{pmatrix}
1 & \dfrac{e}{2\epsilon_F}\\[1mm]
\dfrac{\pi^2}{6}\dfrac{(k_BT)^2}{e\epsilon_F} & 1
\end{pmatrix},
\label{eq:Lambda_z2_2D_app}
\end{equation}
and
\begin{equation}
\Lambda_C^{3D}=\frac{8}{15}v_F\tau_C
\begin{pmatrix}
1 & \dfrac{e}{2\epsilon_F}\\[1mm]
\dfrac{\pi^2}{6}\dfrac{(k_BT)^2}{e\epsilon_F} & 1
\end{pmatrix}.
\label{eq:Lambda_z2_3D_app}
\end{equation}
For both $z=1$ and $z=2$, the explicit dependence on $\tau_R$ cancels from the boundary matrix; the two dispersions differ in that the parabolic band retains thermoelectric boundary couplings whereas the linear band does not.

A partially specular wall is described by the Maxwell mixture of diffuse and specular reflection. If $p$ is the specularity, with $p=0$ for a fully diffuse wall and $p=1$ for a perfectly specular wall, the same half-range construction gives
\begin{equation}
\mathbf J_w=-\Lambda(p)\,\partial_n\mathbf J_w,
\qquad
\Lambda(p)=\frac{1+p}{1-p}\,\Lambda(0).
\label{eq:Lambda_specularity_app}
\end{equation}
The factor $(1+p)/(1-p)$ is the standard specularity renormalization of the boundary length~\cite{Ziman1963}. The diffuse result above is recovered at $p=0$, while the divergence of the slip length as $p\to1$ expresses the absence of tangential momentum transfer at a perfectly specular wall.

The construction also extends locally to a smooth curved surface. At each point one defines $v_n=\vec v\cdot\hat{\mathbf n}(\mathbf r)$ with respect to the local surface normal and applies the same half-range condition in the local tangent plane. The planar result is the leading boundary condition when the local radius of curvature $R_c$ is large compared with the relevant kinetic lengths; curvature produces corrections of relative order $\ell/R_c$ beyond this local planar limit.

\subsection{Boundary reciprocity}

The boundary condition must satisfy the same reciprocity structure as the bulk differential operator. With
\begin{equation}
H=L^{-1}M,
\end{equation}
the required condition is
\begin{equation}
\Lambda^T H=H\Lambda.
\label{eq:boundary_onsager_condition}
\end{equation}

To see why this is sufficient, consider two solutions $\mathbf J^{(a)}$ and $\mathbf J^{(b)}$ driven by uniform forces $\mathbf X^{(a)}$ and $\mathbf X^{(b)}$. Multiplying
\begin{equation}
L^{-1}\mathbf J=\mathbf X+H\nabla^2\mathbf J
\label{eq:reciprocal_diff_eq}
\end{equation}
for solution $b$ by $(\mathbf J^{(a)})^T$, subtracting the equation with $a$ and $b$ interchanged, and integrating over the sample volume gives, because $L^{-1}$ and $H$ are symmetric,
\begin{equation}
\begin{aligned}
\mathbf X^{(b)T}\!\int \mathbf J^{(a)}\,dV
-\mathbf X^{(a)T}\!\int \mathbf J^{(b)}\,dV
={}&\\
-\int_{\partial V}\!
\Big[
\mathbf J^{(a)T}H\,\partial_n\mathbf J^{(b)}
-\mathbf J^{(b)T}H\,\partial_n\mathbf J^{(a)}
\Big]\,dS .
\end{aligned}
\label{eq:D17}
\end{equation}
Using $\mathbf J=-\Lambda\partial_n\mathbf J$ at the wall, the surface term vanishes precisely when Eq.~\eqref{eq:boundary_onsager_condition} is satisfied. Dividing by the sample volume then gives
\begin{equation}
\mathbf X^{(b)T}\overline{\mathbf J}^{(a)}
=
\mathbf X^{(a)T}\overline{\mathbf J}^{(b)}.
\end{equation}
Substituting the definition of the geometry-dependent effective response, $\overline{\mathbf J}=L_{\rm eff}\mathbf X$, and using the fact that the two driving forces are arbitrary, one obtains
\begin{equation}
L_{\rm eff}^T=L_{\rm eff}.
\label{eq:Leff_onsager}
\end{equation}
In particular, $[L_{\rm eff}]_{12}=[L_{\rm eff}]_{21}$ and the confined Seebeck and Peltier responses satisfy the Kelvin relation $\Pi=TS$.

We now verify Eq.~\eqref{eq:boundary_onsager_condition} for the kinetic boundary matrices derived above. For $z=1$, $\Lambda_C=\Lambda_d I$, and therefore $\Lambda_C^T H=H\Lambda_C$ follows immediately. For $z=2$, the ratio $p/v$ is constant, so that the action of the gray Callaway operator closes exactly on the retained current subspace, $RA=AS$. Defining
\begin{equation}
B=A^TA,\qquad
K_0=[A^T|V_n|A]_-,\qquad
K_2=A^TV_n^2A,
\end{equation}
one has $L=BS$ and $W_0=K_0S$. Since the integrand defining $K_2$ is even in $v_n$, the outgoing half-space gives $W_1=K_2S/2$. The kinetic wall matrix and the bulk metric therefore take the forms
\begin{equation}
\Lambda_C=\frac{\tau_C}{2}B K_0^{-1}K_2B^{-1},
\qquad
H=\tau_C B^{-1}K_2B^{-1}.
\label{eq:boundary_onsager_z2_structural}
\end{equation}
Because $B$, $K_0$, and $K_2$ are symmetric,
\begin{equation}
\Lambda_C^T H
=
H\Lambda_C
=
\frac{\tau_C^2}{2}
B^{-1}K_2K_0^{-1}K_2B^{-1}.
\label{eq:boundary_onsager_z2_check}
\end{equation}
Thus the boundary reciprocity condition is exact for the parabolic band within the gray Callaway model; no Sommerfeld truncation is required for this result.

For the partially specular boundary above, $\Lambda(p)$ differs from the diffuse-wall matrix only by a scalar factor, so Eq.~\eqref{eq:boundary_onsager_condition} and the reciprocity of the confined response are unchanged.

\section{Finite-width thermal benchmark}
\label{app:thermal_benchmark}

For completeness, this Appendix records the thermal benchmark discussed earlier in the paper. We first state the exact diffuse-wall result for the relaxation-time Boltzmann equation and the corresponding single-flux FDF solution, and then connect that result with the full two-flux open-circuit problem.

Let
\begin{equation}
w=\frac{W}{v_F\tau_C},
\qquad
F_d(w)=\frac{\kappa_{\rm BTE}(W)}{\kappa_{\rm bulk}}.
\end{equation}
For an energy-independent relaxation time and fully diffuse walls, direct integration of the stationary BTE along trajectories gives in two dimensions
\begin{equation}
F_2(w)=\frac{4}{\pi}\int_0^{\pi/2}\!d\theta\,\cos^2\theta
\left[1-\frac{\sin\theta}{w}
\left(1-e^{-w/\sin\theta}\right)\right],
\label{eq:thermal_BTE_2D}
\end{equation}
and in three dimensions
\begin{equation}
F_3(w)=\frac{3}{2}\int_0^1\!d\mu\,(1-\mu^2)
\left[1-\frac{\mu}{w}
\left(1-e^{-w/\mu}\right)\right].
\label{eq:thermal_BTE_3D}
\end{equation}
These expressions give the exact diffuse-wall BTE benchmark used for the thermal comparison.

The single-flux thermal FDF in the same resistive limit is
\begin{equation}
q_x(y)-\ell_q^2 q_x''(y)=q_{\rm bulk},
\qquad
\ell_q=\frac{v_F\tau_C}{\sqrt{d+2}},
\label{eq:thermal_single_FDF}
\end{equation}
with the diffuse-wall boundary condition
\begin{equation}
q_x(\pm W/2)\pm\ell_s q_x'(\pm W/2)=0.
\label{eq:thermal_single_BC}
\end{equation}
Averaging the solution across the strip gives
\begin{equation}
\frac{\kappa_{\rm FDF}(W)}{\kappa_{\rm bulk}}
=1-\frac{2\ell_q}{W}
\frac{\sinh(W/2\ell_q)}
{\cosh(W/2\ell_q)+(\ell_s/\ell_q)\sinh(W/2\ell_q)}.
\label{eq:thermal_single_ratio}
\end{equation}
For a fully diffuse wall, the half-range boundary construction of Appendix~\ref{app:boundary_conditions} gives
\begin{subequations}
\begin{align}
\ell_s^{(2D)}&=\frac{3\pi}{16}v_F\tau_C,\\
\ell_s^{(3D)}&=\frac{8}{15}v_F\tau_C.
\end{align}
\end{subequations}
For compactness, define $x=\sqrt5\,w/2$. Hence
\begin{subequations}\label{eq:thermal_FDF_explicit}
\begin{align}
\frac{\kappa_{\rm FDF}^{(2D)}}{\kappa_{\rm bulk}}
&=1-\frac{1}{w}
\frac{\sinh w}{\cosh w+(3\pi/8)\sinh w},\\
\frac{\kappa_{\rm FDF}^{(3D)}}{\kappa_{\rm bulk}}
&=1-\frac{2}{\sqrt5\,w}
\frac{\sinh x}{\cosh x+(8\sqrt5/15)\sinh x}.
\end{align}
\end{subequations}
Equations~\eqref{eq:thermal_BTE_2D}--\eqref{eq:thermal_FDF_explicit} provide the corresponding BTE and FDF thermal benchmark.

Figure~\ref{fig:thermal_bte_fdf} compares these two results over the
finite-width crossover. The FDF approaches the spatially resolved BTE
systematically as $W/(v_F\tau_C)$ increases. At
$W=v_F\tau_C$, the relative deviations are approximately $8.4\%$ in
$d=2$ and $7.6\%$ in $d=3$, and decrease rapidly for larger widths.
The discrepancy is largest in the strongly kinetic regime
$W<v_F\tau_C$, where a second-order spatial closure is not expected
to reproduce the detailed boundary-layer dynamics. The comparison
therefore does not define a sharp lower cutoff for the constitutive
description, but shows that the integrated thermal response remains
quantitatively useful down to widths of order $v_F\tau_C$.

{
\begin{figure}[t]
    \centering
    \IfFileExists{Fig5_JT.pdf}{\includegraphics[width=0.9\columnwidth]{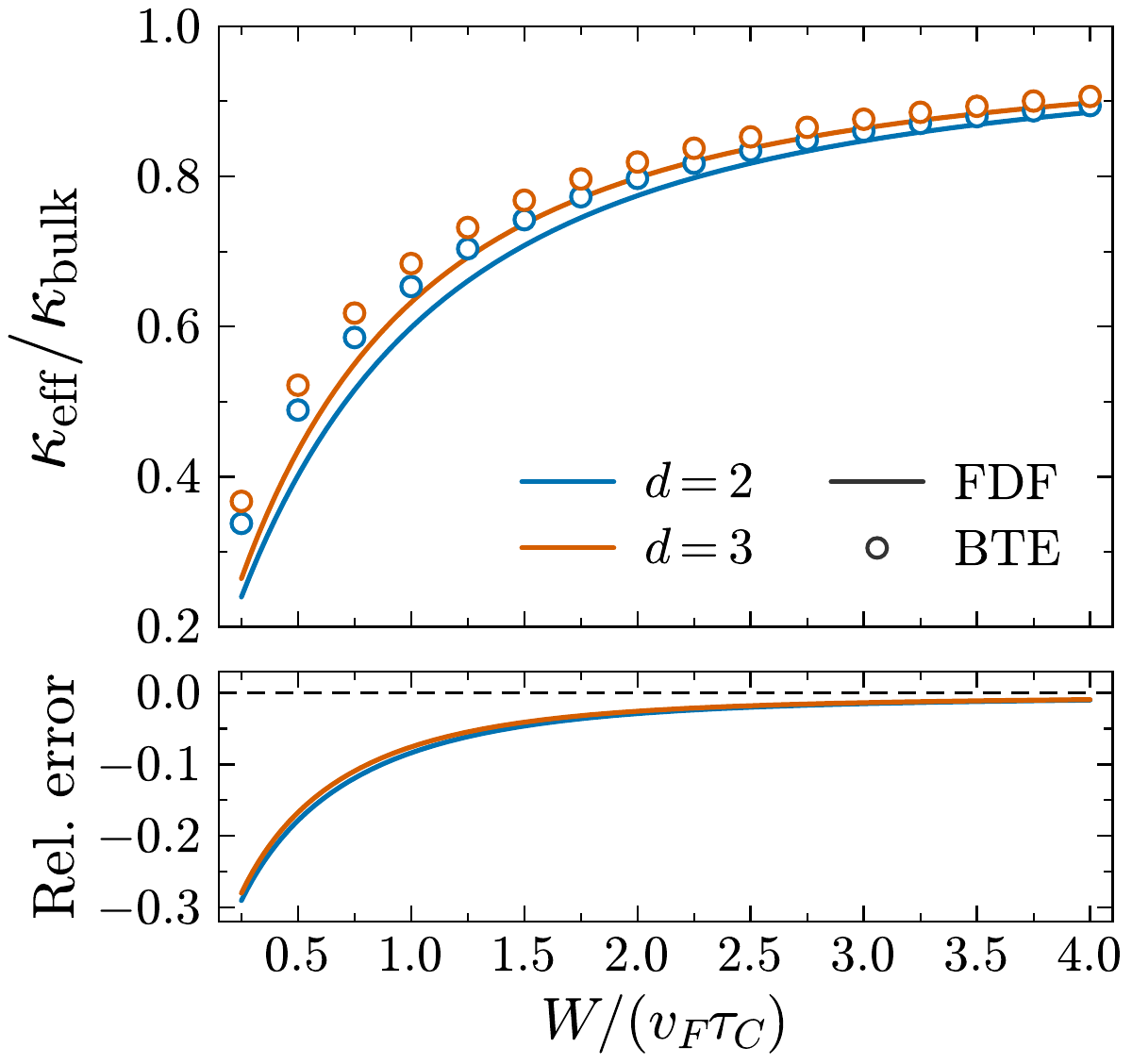}}{}
{\color{black}
    \caption{Finite-width thermal transport in the resistive
    limit, $r=1$. Normalized open-circuit thermal conductivity as a function
    of $W/(v_F\tau_C)$ for $d=2$ and $d=3$. Solid lines show the FDF
    result and open circles the exact diffuse-wall BTE result for the
    same relaxation-time model. The agreement with the BTE improves
    systematically as the thickness exceeds the fast kinetic propagation length.}
    \label{fig:thermal_bte_fdf}
}
\end{figure}
}

{\color{black}
As discussed in Sec.~VI, for $r=1$ the thermoelectric feedback is beyond
the leading degenerate thermal response.  This explains why the single-flux
FDF provides the appropriate leading-order description for the benchmark in
Fig.~\ref{fig:thermal_bte_fdf}.  In the next subsection, we establish the
corresponding result directly from the coupled FDF for arbitrary
$0<r\leq1$.
}

\subsection{Open-circuit thermal transport in the coupled FDF}
\label{app:coupled_open_thermal}

For arbitrary $0<r\leq1$, the exact local heat equation obtained after
eliminating the electric field is
\begin{equation}
\begin{split}
\mathbf q={}&-\kappa_{\rm oc}\nabla T
+\Pi_0\mathbf j
+\mathcal M_j\nabla^2\mathbf j
+\mathcal M_q\nabla^2\mathbf q,
\end{split}
\label{eq:heat_twoflux_app}
\end{equation}
where
\begin{equation}
\begin{aligned}
\Pi_0&=\frac{L_{21}}{L_{11}},\\
\mathcal M_j&=M_{21}-\Pi_0M_{11},\qquad
\mathcal M_q=M_{22}-\Pi_0M_{12}.
\end{aligned}
\label{eq:Mphysical_app}
\end{equation}
and
\begin{equation}
\kappa_{\rm oc}=\frac1T\left(L_{22}-\frac{L_{21}L_{12}}{L_{11}}\right).
\end{equation}
The reduction must be made on the combinations in
Eq.~\eqref{eq:Mphysical_app}, rather than by assigning orders to the
individual entries of $M$.  In particular, because the heat-sector
moments begin at $O(t_2)$, the $O(t_2^2)$ Sommerfeld moments in
Eq.~\eqref{eq:Sommerfeld_app} contribute to the relative $O(t_2/\mu^2)$
correction to $\mathcal M_q$.

For $\epsilon=Ap^z$, direct expansion of the complete matrices $B$, $L$,
and $D$ before inversion gives
\begin{subequations}\label{eq:open_combinations_app}
\begin{align}
\Pi_0&=\frac{t_2}{e\mu}\,
\frac{d+(z-2)r}{z}+O(T^4),\\
\mathcal M_j&=\ell_C^2\frac{t_2}{e\mu}\,
\frac{z-d}{z}+O(T^4),\\
\mathcal M_q&=\ell_C^2\left[1+
C_q(z,d,r)\frac{t_2}{\mu^2}
+O\!\left(\frac{t_2^2}{\mu^4}\right)\right],
\end{align}
\end{subequations}
with
\begin{equation}
\begin{split}
C_q(z,d,r)={}&
\frac{22}{5}\frac{(d+z-2)(z-1)}{z^2}
+(1-r)\frac{(2-z)^2}{z^2}\\
&+\frac{21}{5}\frac{(z-1)(z-2)}{z^2}.
\end{split}
\label{eq:Cq_app}
\end{equation}
The important point is not the particular form of $C_q$, but that it is
finite as $r\to0$: all apparent $1/r$ enhancements cancel after the
physical open-circuit combinations are formed.

It remains to verify that the local counterflow does not acquire an inverse
power of $r$.  In a strip, subtracting the cross-section average of the
charge equation from the local equation removes the uniform open-circuit
field.  For any nonuniform transverse mode with wave number $k$ one obtains,
to leading degenerate order,
\begin{equation}
(1+k^2M_{11})j_k=-k^2M_{12}q_k,
\label{eq:jk_qk_app}
\end{equation}
so that, using the leading power-law coefficients,
\begin{equation}
\frac{k^2M_{12}}{1+k^2M_{11}}
=\frac{e}{z\mu}\,
\frac{k^2\ell_C^2\,[z+r(z-2)]}{r+k^2\ell_C^2}.
\label{eq:counterflow_regular_app}
\end{equation}
For every nonuniform mode of a fixed finite geometry this expression is
bounded and tends to $e/\mu$ as $r\to0$.  The kinetic boundary matrix has
the same regular Sommerfeld expansion, so the diffuse-wall boundary
condition introduces no inverse thermoelectric factor.  Thus the
open-circuit profile $j(y)$ remains first order in the thermoelectric
coupling; the condition $\overline j=0$ removes only its uniform component
and does not require $j(y)$ to vanish pointwise.

{\color{black}
Equations~\eqref{eq:open_combinations_app}--\eqref{eq:counterflow_regular_app}
show that all thermoelectric feedback terms are beyond the leading degenerate
thermal response.  Therefore, for arbitrary $0<r\leq1$, the coupled
open-circuit FDF gives at leading order
\begin{equation}
\mathbf q=-\kappa_{\rm oc}\nabla T+\ell_C^2\nabla^2\mathbf q,
\qquad
\ell_C^2=\frac{v_F^2\tau_C^2}{d+2}.
\label{eq:heat_GK_app_final}
\end{equation}
This is Eq.~\eqref{eq:heat_GK_main} of the main text and establishes why the
single-flux FDF used in Fig.~\ref{fig:thermal_bte_fdf} gives the leading
coupled open-circuit thermal response.}

{
Since both the propagation length $\ell_C$ and the leading diffuse-wall slip
length are proportional to $v_F\tau_C$ and independent of $r$, the normalized
FDF curve in Fig.~\ref{fig:thermal_bte_fdf}, plotted against
$W/(v_F\tau_C)$, is the leading-degenerate prediction for arbitrary
$0<r\leq1$. The exact BTE symbols shown in that figure, however, correspond
specifically to the resistive case $r=1$.
}

\section{Finite-width Seebeck benchmark}
\label{app:seebeck_benchmark}

This Appendix gives the two calculations used in the quantitative comparison of Sec.~\ref{sec:strip_BTE_benchmark}. We first derive the finite-width Seebeck coefficient from the full coupled FDF boundary-value problem and then summarize the spatially resolved Callaway Boltzmann calculation. In both descriptions open circuit is imposed only on the cross-section-averaged electric current, so local thermoelectric counterflows are retained.

\subsection{FDF finite-width Seebeck coefficient}
\label{app:FDF_seebeck}

For a strip of width $W$, infinite along $x$, the longitudinal charge and heat currents are collected in
\begin{equation}
 \mathbf J(y)=\begin{pmatrix}j(y)\\ q(y)\end{pmatrix},
 \qquad
 \mathbf X=\begin{pmatrix}E\\ -\nabla T/T\end{pmatrix},
\end{equation}
and the coupled FDF equation is
\begin{equation}
 \mathbf J(y)=L\mathbf X+M\mathbf J''(y).
 \label{eq:app_FDF_strip}
\end{equation}
The bulk solution is $\mathbf J_0=L\mathbf X$. Defining
\begin{equation}
 \delta\mathbf J(y)=\mathbf J(y)-\mathbf J_0,
\end{equation}
Eq.~\eqref{eq:app_FDF_strip} becomes
\begin{equation}
 \delta\mathbf J=M\,\delta\mathbf J''.
 \label{eq:app_homogeneous}
\end{equation}
Let
\begin{equation}
M=V D_\lambda^2 V^{-1},
\qquad
D_\lambda=\operatorname{diag}(\lambda_+,\lambda_-),
\tag{G5}
\end{equation}
where the columns of $V$ are the right eigenvectors of $M$,
and $\lambda_\pm^2$ are its eigenvalues, defining the two
nonlocal lengths $\lambda_\pm$. For identical walls the physical solution is even in $y$ and can be written
\begin{equation}
 \mathbf J(y)=\mathbf J_0+V\cosh(D_\lambda^{-1}y)\,\mathbf C,
\label{eq:app_modal_solution}
\end{equation}
 with $\mathbf C$ to be determined by boundary conditions. At the upper wall, $y=W/2$, define
\begin{equation}
 D_c=\cosh\!\left(\frac{W}{2}D_\lambda^{-1}\right),
 \qquad
 D_s=D_\lambda^{-1}\sinh\!\left(\frac{W}{2}D_\lambda^{-1}\right),
\end{equation}
and $\widetilde\Lambda=V^{-1}\Lambda V$. The kinetic boundary condition derived in Appendix~\ref{app:boundary_conditions},
\begin{equation}
 \mathbf J(W/2)=-\Lambda\mathbf J'(W/2),
 \label{eq:app_robin}
\end{equation}
gives
\begin{equation}
 \mathbf C
 =-\left(D_c+\widetilde\Lambda D_s\right)^{-1}V^{-1}\mathbf J_0.
 \label{eq:app_C}
\end{equation}
The lower-wall condition is then automatically satisfied by the even solution.

The transverse average is
\begin{equation}
 \overline{\mathbf J}
 =\frac{1}{W}\int_{-W/2}^{W/2}\mathbf J(y)\,dy
 =\mathbf J_0+V D_a\mathbf C,
\end{equation}
with
\begin{equation}
 D_a=\frac{2}{W}D_\lambda
 \sinh\!\left(\frac{W}{2}D_\lambda^{-1}\right).
\end{equation}
Using Eq.~\eqref{eq:app_C},
\begin{equation}
 \overline{\mathbf J}=L_{\rm eff}(W)\mathbf X,
 \label{eq:app_Leff_def}
\end{equation}
where
\begin{equation}
 L_{\rm eff}(W)=
 \left[I-V D_a
 \left(D_c+\widetilde\Lambda D_s\right)^{-1}V^{-1}\right]L .
 \label{eq:app_Leff}
\end{equation}
This is the full two-flux FDF response used for the finite-width benchmark. No pointwise condition $j(y)=0$ and no reduction to a single heat-current equation is made.

Writing Eq.~\eqref{eq:app_Leff_def} componentwise,
\begin{equation}
 \overline j
 =[L_{\rm eff}]_{11}E
 -[L_{\rm eff}]_{12}\frac{\nabla T}{T}.
\end{equation}
The physical open-circuit condition is
\begin{equation}
 \overline j=0,
 \label{eq:app_open_circuit}
\end{equation}
which fixes the longitudinal electric field after the local current profile has been solved. Hence
\begin{equation}
 S_{\rm FDF}(W)
 =\frac{E}{\nabla T}
 =\frac{[L_{\rm eff}(W)]_{12}}
 {T[L_{\rm eff}(W)]_{11}} .
 \label{eq:app_SFDF}
\end{equation}
With $S_0=L_{12}/(T L_{11})$, the normalized quantity plotted in the benchmark is therefore
\begin{equation}
 \frac{S_{\rm FDF}(W)}{S_0}
 =\frac{[L_{\rm eff}(W)]_{12}/[L_{\rm eff}(W)]_{11}}
 {L_{12}/L_{11}} .
 \label{eq:app_SFDF_normalized}
\end{equation}
Equations~\eqref{eq:app_Leff} and \eqref{eq:app_SFDF_normalized}, evaluated with the Callaway matrices $L$, $M$, and the kinetic wall matrix $\Lambda$, generate the FDF curves in the Seebeck benchmark.

When the wall matrix is scalar, $\Lambda=\Lambda_B I$, the normal modes decouple. Each modal amplitude $g_\alpha$ obeys
\begin{equation}
 g_\alpha(y)=g_{\alpha0}
 +C_\alpha\cosh(y/\lambda_\alpha),
\end{equation}
with
\begin{equation}
 C_\alpha=-\frac{g_{\alpha0}}
 {\cosh h_\alpha+(\Lambda_B/\lambda_\alpha)\sinh h_\alpha},
 \qquad
 h_\alpha=\frac{W}{2\lambda_\alpha}.
\end{equation}
Averaging gives $\overline g_\alpha=f_\alpha g_{\alpha0}$, where
\begin{equation}
 f_\alpha=1-\frac{2\lambda_\alpha}{W}
 \frac{\sinh h_\alpha}
 {\cosh h_\alpha+(\Lambda_B/\lambda_\alpha)\sinh h_\alpha}.
 \label{eq:app_modal_factor}
\end{equation}
For a linear band ($z=1$), the diffuse Callaway wall matrix is proportional to the identity, so Eq.~\eqref{eq:app_modal_factor} provides the two suppression factors associated with the momentum and orthogonal modes. Direct substitution into the Sommerfeld form of $L$ gives the analytic finite-width Seebeck result quoted in the main text. In particular, for $r=1$ the two modes have the same propagation and boundary factors, which cancel in the ratio in Eq.~\eqref{eq:app_SFDF_normalized}, giving $S_{\rm FDF}/S_0=1$. For a parabolic band ($z=2$), the first Sommerfeld corrections make $V^{-1}\Lambda V$ generally non-diagonal; the full matrix expression \eqref{eq:app_Leff} is therefore retained in the numerical evaluation.

\subsection{Boltzmann-equation film benchmark}
\label{app:BTE}

\begin{figure*}[t!]
\centering
\includegraphics[width=0.88\textwidth]{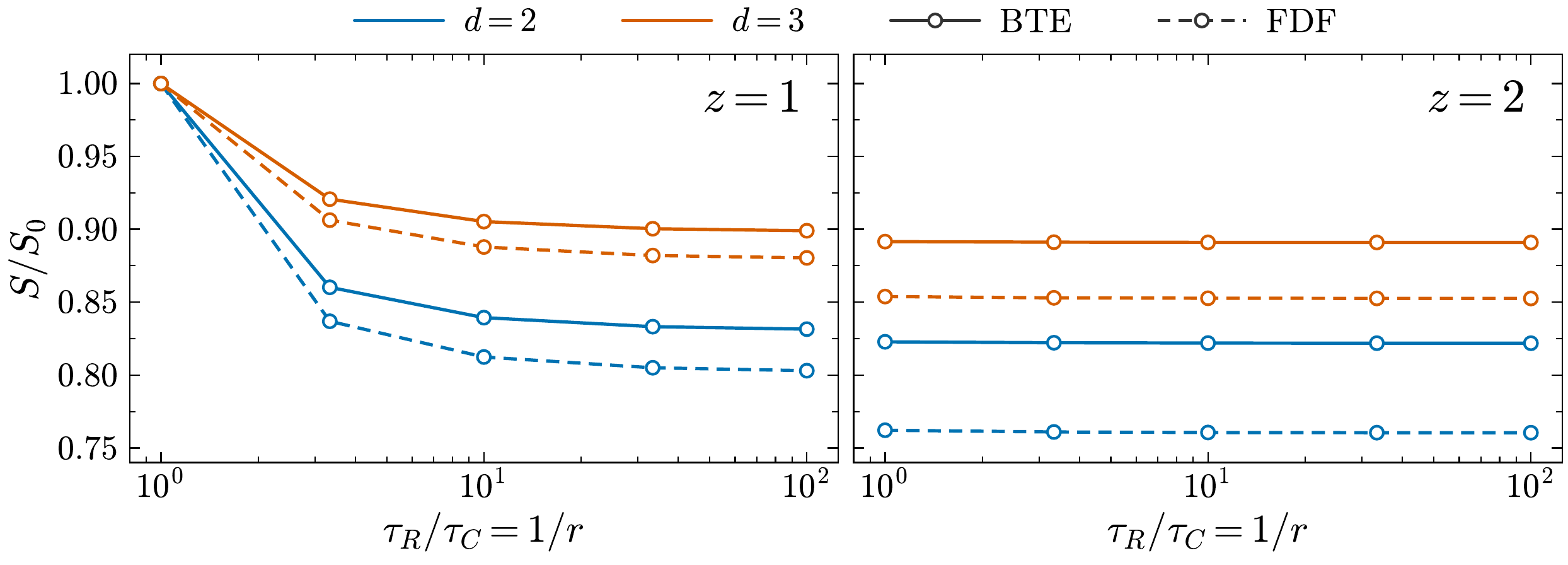}
\caption{Systematic finite-width benchmark of the normalized Seebeck coefficient for linear ($z=1$, left) and parabolic ($z=2$, right) dispersions in $d=2$ and $d=3$. Lines connect the FDF and spatially resolved BTE results as the relative strength of momentum-conserving and momentum-relaxing scattering is varied. The calculation uses the common thickness $W=v_F\tau_C$. The comparison confirms that the second-order closure remains quantitatively useful at widths of order the fast kinetic propagation length and becomes more accurate as the width is increased, as shown in Fig.~\ref{fig:widthscan_main}.}
\label{fig:BTEFDF_app}
\end{figure*}

For the quantitative comparison with Eq.~\eqref{eq:app_SFDF_normalized}, we solve the Callaway BTE in the same strip with energy-independent relaxation times,
\begin{equation}
 v_y\partial_y\phi+\frac{\phi}{\tau_C}
 =A\mathbf X+\frac{1-r}{\tau_C}p_xu(y),
 \label{eq:app_BTE}
\end{equation}
with the local momentum self-consistency condition
\begin{equation}
 u(y)=\frac{\langle p_x\phi(y)\rangle}
 {\langle p_x^2\rangle}.
 \label{eq:app_self}
\end{equation}
Diffuse incoming boundary conditions are imposed trajectory by trajectory. Integration of Eq.~\eqref{eq:app_BTE} along characteristics expresses $\phi$ as a direct wall-to-point contribution plus an integral over $u(y')$. Substitution into Eq.~\eqref{eq:app_self} gives a Fredholm equation of the form
\begin{equation}
 [I-(1-r)\mathcal G]u=u^{(0)}.
 \label{eq:app_Fredholm}
\end{equation}
The equation is solved independently for electrical and thermal forcing. The distribution is then reconstructed, the local charge and heat currents are formed, and only at the final stage are they averaged across the strip,
\begin{equation}
 \overline{\mathbf J}=L_{\rm BTE}(W,r)\mathbf X.
 \label{eq:app_LBTE}
\end{equation}
Accordingly, the BTE Seebeck coefficient is extracted from the same averaged open-circuit prescription as in the FDF,
\begin{equation}
 S_{\rm BTE}(W)
 =\frac{[L_{\rm BTE}(W,r)]_{12}}
 {T[L_{\rm BTE}(W,r)]_{11}} .
 \label{eq:app_SBTE}
\end{equation}
Thus the FDF--BTE comparison uses identical macroscopic forcing and the same physical open-circuit constraint; neither calculation suppresses the local countercurrents by imposing $j(y)=0$ pointwise. In the BTE, in addition, the collective amplitude $u(y)$ is determined self-consistently and is not assumed spatially uniform.

The systematic benchmark is performed for $z=1,2$ and $d=2,3$ as the ratio $\tau_R/\tau_C=1/r$ is varied. At the common reference width $W=v_F\tau_C$, the linear-band FDF--BTE discrepancy remains within about $3.4\%$ in two dimensions and $2.1\%$ in three dimensions over the range considered. For the parabolic band the corresponding maximum discrepancies are about $7.5\%$ and $4.3\%$. These fixed-width results complement the width scan in the main text, which shows directly that the agreement improves as $W/(v_F\tau_C)$ increases.

The qualitatively different trends for linear and parabolic bands can be traced to the energy dependence of the bounded kinetic problem. For an isotropic power-law band $\epsilon=Ap^z$,
\begin{equation}
 v_x=\alpha(\epsilon)p_x,
 \qquad
 \alpha(\epsilon)=zAp^{z-2}.
\end{equation}
If the complete bounded propagator is energy independent, the direct-current and momentum-projected channels carry, in the degenerate limit, spectral powers
\begin{equation}
 n_2=\frac{d+z-2}{z},
 \qquad
 n_1=\frac{d}{z},
 \qquad
 n_2-n_1=\frac{z-2}{z}.
\end{equation}
For $z=2$, $v_x=p_x/m$ and the two energy moments coincide. An energy-independent bounded propagator would therefore leave the normalized Seebeck coefficient unchanged. The physical parabolic problem does not satisfy that premise because $v_y(\epsilon,\theta)\propto\sqrt{\epsilon}\sin\theta$; its finite-width thermopower is consequently produced by boundary energy filtering. For $z=1$, instead, $v=v_F$ makes the bounded propagation energy independent, but $v_x=v_Fp_x/p$ is not proportional to $p_x$. The Callaway momentum channel then changes the thermoelectric ratio for $r<1$, producing the collective finite-width correction seen in Figs.~\ref{fig:widthscan_main} and \ref{fig:BTEFDF_app}.

As a complementary view of the same benchmark, Fig.~\ref{fig:BTEFDF_app} shows the dependence on the collision-time ratio $r$ at the fixed width $W=v_F\tau_C$, for $z=1,2$ and $d=2,3$. The FDF and BTE quantities are evaluated with the same averaged open-circuit prescription used above; no additional approximation is introduced.

\section{Open-circuit counterflow and finite-sample closure}
\label{app:counterflow_closure}

This Appendix clarifies why open circuit constrains the net electrical current rather than the local current and describes a finite-rectangle calculation illustrating the closure of electrical counterflows.

Starting from the charge constitutive equation,
\begin{equation}
\mathbf j=L_{11}\mathbf E-\frac{L_{12}}{T}\nabla T
+M_{11}\nabla^2\mathbf j+M_{12}\nabla^2\mathbf q,
\label{eq:counterflow_charge_app}
\end{equation}
suppose, more restrictively, that $\mathbf j(\mathbf r)=0$ everywhere. The electrochemical field would then satisfy
\begin{equation}
\mathbf E=S_0\nabla T-\frac{M_{12}}{L_{11}}\nabla^2\mathbf q.
\label{eq:E_pointwise_zeroj_app}
\end{equation}
Because $\nabla\times\mathbf E=0$, pointwise cancellation requires
\begin{equation}
M_{12}\nabla\times\nabla^2\mathbf q=0.
\label{eq:compatibility_zeroj_app}
\end{equation}
For $\mathbf q=q_x(y)\hat{\mathbf x}$ this reduces to
\begin{equation}
M_{12}q_x'''(y)=0.
\label{eq:qthird_zeroj_app}
\end{equation}
A diffuse-wall thermal profile generically violates this condition when $M_{12}\ne0$. Thus the pointwise constraint $\mathbf j=0$ is generally incompatible with a conservative electrochemical field. Exceptional cases include $M_{12}=0$ and special heat-current profiles for which $\nabla\times\nabla^2\mathbf q=0$. The same distinction is apparent in the spatially resolved Boltzmann benchmark of Appendix~\ref{app:seebeck_benchmark}: electric and thermal driving profiles are not generally proportional across the strip.

In a finite electrically insulating sample, charge conservation and impermeability,
\begin{equation}
\nabla\cdot\mathbf j=0,\qquad \mathbf j\cdot\hat{\mathbf n}=0,
\label{eq:finite_closure_constraints}
\end{equation}
require oppositely directed longitudinal currents to turn through transverse currents near the ends. We illustrate this closure using the coupled FDF equations in a rectangle $0<x<L_x$, $-W/2<y<W/2$, with $L_x=6W$, $W=v_F\tau_C$, $d=2$, $z=2$, and $r=\tau_C/\tau_R=1$. At the retained Sommerfeld order, the nonlocal and diffuse-wall slip matrices are
\begin{align}
M&=\frac{W^2}{4}
\begin{pmatrix}
1&e/\epsilon_F\\
\dfrac{\pi^2(k_BT)^2}{6e\epsilon_F}&1
\end{pmatrix},\label{eq:counterflow_M_app}\\
\Lambda_C&=\frac{3\pi W}{16}
\begin{pmatrix}
1&\dfrac{e}{2\epsilon_F}\\
\dfrac{\pi^2(k_BT)^2}{6e\epsilon_F}&1
\end{pmatrix}.
\label{eq:counterflow_matrices_app}
\end{align}
The sidewalls obey $j_y=q_y=0$ and the matrix tangential-slip condition. At the two ends we impose $j_x=0$, uniform temperatures $T_1>T_2$, and the same matrix slip condition for the tangential components $(j_y,q_y)$, as an idealized thermal-contact model. In particular, $q_x$ is not constrained to vanish at either end.

For the divergence-free two-dimensional currents, introduce stream functions
\begin{align}
\mathbf j&=(\partial_y\psi,-\partial_x\psi),
&\mathbf q&=(\partial_y\chi,-\partial_x\chi),
\label{eq:counterflow_stream_app}\\
U&=\nabla^2\psi,&V&=\nabla^2\chi.
\label{eq:counterflow_vorticities_app}
\end{align}
Charge and energy conservation then hold identically. Taking the curl of the constitutive equations for $\mathbf j$ and $\mathbf q$ gives
\begin{align}
U-M_{11}\nabla^2U-M_{12}\nabla^2V&=0,\\
V-M_{21}\nabla^2U-M_{22}\nabla^2V&=0.
\label{eq:counterflow_curl_app}
\end{align}
These equations, the definitions of $U,V$, and the boundary conditions define the finite-geometry problem. The isothermal-contact condition fixes the tangential thermal force; the overall temperature difference fixes the linear-response amplitude, without setting the contact heat flux to zero.

A finite-difference solution for $k_BT/\epsilon_F=10^{-3}$,
normalized by the net heat current, gives the electrical-current
streamlines shown in Fig.~\ref{fig:counterflow_loops_app}.
The longitudinal current reverses direction across the strip,
with counterflows closing through transverse currents near the
insulating ends. 

\bibliographystyle{apsrev4-2}
\bibliography{library_v17}

\end{document}